\documentclass[a4paper,spanish,english,onecolumn,aps,pre,superscriptaddress]{revtex4-2}
\usepackage[T1]{fontenc}
\usepackage[latin9]{inputenc}
\usepackage{color}
\usepackage{float}
\usepackage{textcomp}
\usepackage{amsmath}
\usepackage{graphicx}

\makeatletter

\pdfpageheight\paperheight
\pdfpagewidth\paperwidth

\providecommand{\tabularnewline}{\\}
\floatstyle{ruled}
\newfloat{algorithm}{H}{loa}
\providecommand{\algorithmname}{Algorithm}
\floatname{algorithm}{\protect\algorithmname}

\@ifundefined{date}{}{\date{}}

\makeatother

\usepackage{babel}
\addto\shorthandsspanish{\spanishdeactivate{~<>}}

\addto\captionsspanish{\renewcommand{\algorithmname}{Algoritmo}}

\begin{document}
\selectlanguage{spanish}%
\begin{center}
\textbf{\large{}Representación de redes arquimedianas e inclusión:
aplicaciones computacionales en percolación y transiciones en redes}{\large\par}
\par\end{center}
\author{Auro Anibal Torres}
\author{Antonio Jose Ramirez-Pastor}
\affiliation{Departamento de Física, Instituto de Física Aplicada (INFAP), Universidad
Nacional de San Luis--CONICET}
\address{Ejército de Los Andes 950, D5700HHW San Luis, Argentina}
\email{auro.torres@gmail.com}

\date{Julio de 2025}
\begin{abstract}
Presentamos una representación geométrica alternativa para las once
redes Arquimedianas, en la que cada sitio y enlace se etiqueta de
forma única mediante un par ordenado de enteros y se caracteriza mediante
la función módulo. Este esquema de etiquetado y caracterización estructurado
permite implementar modelos computacionales de manera eficiente y
sistemática sobre estas redes, sin recurrir a indexaciones ad hoc,
y ofrece un marco versátil para futuros estudios sobre teselaciones
regulares. Como aplicación obtenemos para cada una de las redes Arquimedianas,
las curvas de fase por deposición de monómeros para los modelos de
percolación sitio-enlace conocidos en la literatura como $S\cup B$
y $S\cap B$. Mostramos que las curvas de fase se ordenan en el espacio
de fase según lo inducen las relaciones de inclusión parcial y total
entre ellas demostradas por Parviainen et al (2003). Además, para
los pares de redes $(3.6.3.6)/(3.4.6.4)$ y $(3^{3}.4^{2})/(3^{2}.4.3.4)$,
observamos una inversión entre los umbrales críticos de percolación
pura de sitios y de enlaces, y mostramos que este fenómeno está vinculado
al cruce entre sus curvas de fase, evidenciando la sensibilidad del
comportamiento crítico frente a la topología de la red.
\end{abstract}
\maketitle

\section{Introducción}

Denotamos con $G$ un grafo o red cualquiera, con $V(G)$ su conjunto
de sitios, y con $E(G)$ su conjunto de enlaces \citep{newman2010networks}.

En un problema clásico de percolación pura de sitios sobre $G$, dos
sitios primeros vecinos cualesquiera de $V(G)$, se consideran conectados
entre si, si ambos sitios han sido ocupados con objetos físicos de
interés, mediante algún proceso de ocupación predefinido, que comúnmente
se elige aleatorio. Un cluster de sitios ocupados, es un conjunto
de sitios primeros vecinos ocupados, simplemente conexo. Es decir,
dado dos sitios cualesquiera del cluster, siempre es posible encontrar
un camino constituido por sitios primeros vecinos ocupados, por el
cual se puede ir de un sitio al otro. Se considera que la red ha percolado
en una dada dirección, cuando un cluster de sitios ocupados, se extiende
de borde a borde de la red, en la dirección elegida.

Análogamente, en un problema clásico de percolación pura de enlaces
sobre $G$, dos enlaces primeros vecinos cualesquiera de $E(G)$,
se consideran conectados entre si, si ambos enlaces han sido ocupados
con objetos físicos de interés, mediante algún proceso de ocupación
predefinido, que comúnmente se elige aleatorio. Un cluster de enlaces
ocupados es un conjunto de enlaces primeros vecinos ocupados, simplemente
conexo. Es decir, dado dos enlaces cualesquiera del cluster, siempre
se encontrar un camino constituido por enlaces primeros vecinos ocupados,
por el cual se puede ir de un enlace al otro. Se considera que la
red ha percolado en una dada dirección, cuando un cluster de enlaces
ocupados, se extiende de borde a borde de la red, en la dirección
elegida. 

Resumiendo, en un problema de percolación pura de sitios solo focalizamos
la atención sobre el conjunto de sitios $V(G)$ de la red, mientras
que en un problema de percolación pura de enlaces, solo focalizamos
la atención sobre su conjunto de enlaces $E(G)$.

\begin{figure}
\begin{minipage}[t]{0.5\columnwidth}%
\begin{center}
\includegraphics[width=7cm]{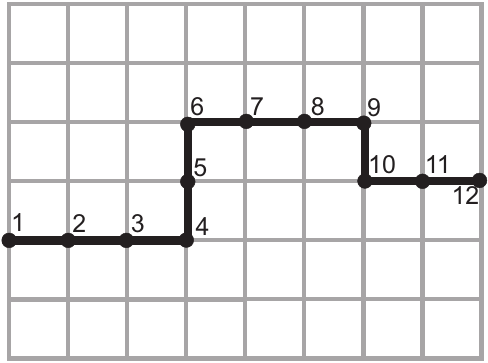}
\par\end{center}
\begin{center}
\textbf{(a)}
\par\end{center}%
\end{minipage}%
\begin{minipage}[t]{0.5\columnwidth}%
\begin{center}
\includegraphics[width=7cm]{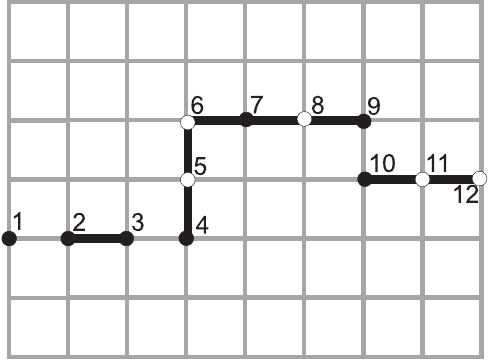}
\par\end{center}
\begin{center}
\textbf{(b)}
\par\end{center}%
\end{minipage}

\caption{\label{fig:Conexionentresitios-modelos-AND-OR-arXiv}Un sitio ocupado
se indica con un circulo lleno en color negro. Un enlace ocupado se
indica con una línea gruesa de color negro.\textbf{ (a)} Conexión
de sitios adyacentes en el modelo $S\cap B$: Cualquier par de sitios
adyacentes $(i,i+1)$ están conectados, pues estos sitios y el enlace
que los une están ocupados. El cluster de sitios y enlaces ocupados
que va del sitio 1 al sitio 10 atraviesa la red de borde a borde,
horizontalmente. decimos que la red ha percolado horizontalmente.
(b) Conexión de sitios adyacentes en el modelo $S\cup B$: Cualquier
par de sitios adyacentes $(i,i+1)$ están conectados, ya sea porque
ambos sitios están ocupados, o el enlace que los une está ocupado,
o ambas cosas a la vez. Como los sitios 1 y 10 están conectados, y
pertenecen a bordes opuestos, decimos que la red ha percolado horizontalmente.}
\end{figure}

Por otra parte, un problema de percolación mixta sitio-enlace se supone
que es una generalización del problema de percolación pura de sitios,
y también el de enlaces. Ahora bien, geométricamente, el primer vecino
de un sitio ya no será otro sitio, sino que será un enlace. Y viceversa,
el primer vecino de un enlace será un sitio. Dado un sitio cualquiera
de un grafo $G$, el sitio vecino más cercano a este será uno de sus
sitios segundos vecinos. Y dado un enlace de $G$, el enlace más cercano
a este será uno de sus enlaces segundos vecinos. Cualquiera sea el
modelo de percolación mixta, este debe redefinir el concepto de conexión
entre pares de sitios segundos vecinos, o bien entre pares de enlaces
segundos vecinos. En la literatura se distinguen al menos dos modelos
de percolación mixta sitio-enlace. Hablaremos aquí solo de los dos
modelos implicados en este trabajo, que son el modelo de percolación
mixta de sitios y enlaces ($S\cap B$) \citep{gonzalez2013site,gonzalez2016site},
y el modelo de percolación mixta de sitios o enlaces ($S\cup B$)
\citep{gonzalez2013site,gonzalez2016site}. En adelante nos referiremos
a ellos simplemente como los modelos $S\cap B$ y $S\cup B$, respectivamente.
A continuación definimos cuando dos sitios segundos vecinos están
conectados en ambos modelos:

\textbullet{} Modelo $S\cap B$ \citep{gonzalez2013site,gonzalez2016site}:
Ver Figura \ref{fig:Conexionentresitios-modelos-AND-OR-arXiv}\textbf{(a)}.
Dos sitios segundos vecinos de la red se consideran conectados entre
sí, si ambos sitios y el enlace que los une están ocupados. Así, dos
sitios distantes de la red se consideran conectados si una secuencia
de sitios segundos vecinos conectados y consecutivos los une. Por
otra parte, un clúster es un conjunto de sitios y enlaces ocupados
en el cual, cualquier par de enlaces segundos vecinos están unidos
entre sí por sitios ocupados, y a la vez, cualquier par de sitios
segundos vecinos son unidos entre si por enlaces ocupados. En este
modelo solo podemos hablar de clústeres de sitios y enlaces ocupados.
En resumen, podemos interpretar a este modelo como una interacción
entre sitios y enlaces, o entre enlaces y sitios, hasta el primer
vecino. La expresión lógica de este modelo podemos expresarla como:
$C_{ij}=s_{i}\wedge e_{ij}\wedge s_{j}$, donde $s_{i}$ es un segundo
vecino de $s_{j}$ ( o viceversa), y $e_{ij}$ es el enlace que los
une. 

\textbullet{} \textbf{Modelo} $\mathbf{S\cup B}$ \citep{gonzalez2013site,gonzalez2016site}:
Ver Figura \ref{fig:Conexionentresitios-modelos-AND-OR-arXiv}\textbf{(b)}.
Dos sitios segundos vecinos de la red se consideran conectados entre
sí, si ambos sitios están ocupados, o bien si el enlace que los une
esta ocupado, o bien cuando se dan ambas situaciones. Así, dos sitios
distantes de la red se consideran conectados si una secuencia de pares
de sitios segundos vecinos conectados y consecutivos los une. En este
modelo, podemos hablar de al menos dos tipos de clústeres: \textbf{1)}\textbf{\textit{
Clúster de sitios y enlaces no necesariamente ocupados}}. Es un conjunto
de sitios y enlaces no necesariamente ocupados, simplemente conexo.
Es decir, se puede ir de un sitio a otro del cluster a través de una
sucesión consecutiva de pares de sitios segundos vecinos conectados.
\textbf{2)} \textbf{\textit{Clúster de sitios y enlaces ocupados}}.
Conformado por un clúster del tipo \textbf{1)} pero en dónde solo
se contabilizan los sitios y enlaces ocupados. En resumen, podemos
considerar a este modelo como una interacción entre sitios y enlaces,
o entre enlaces y sitios, hasta el segundo vecino. La expresión lógica
de este modelo podemos expresarla como: $C_{ij}=(s_{i}\wedge s_{j})\vee e_{ij}$,
donde $s_{i}$ es un segundo vecino de $s_{j}$ ( o viceversa), y
$e_{ij}$ es el enlace que los une. 

En cualquiera de los dos modelos, cuando uno de los sitios distantes
conectados pertenece a uno de los bordes de la red, y el otro pertenece
al borde opuesto de la misma, se dice que la red ha percolado en la
dirección que une ambos bordes. Al cluster que contiene los sitios
conectados de ambos bordes, se le llama cluster gigante si la red
es de tamaño finito, y cluster infinito si la red es de tamaño infinito.
Su presencia determina una \textit{transición de fase geométrica}
en el sistema. 

La curva de fase asociada a la deposición de monómeros sobre $G$,
surge cuando se ocupan aleatoriamente los enlaces de $E(G)$ con probabilidad
$P_{b}$, y luego se ocupan con monómeros los sitios de $V(G)$ con
probabilidad $P_{s}$, hasta encontrar el menor valor de $P_{s}$
que haga percolar la red en alguna dirección, para un dado $P_{b}$.
Esto determina una probabilidad crítica $P_{s,c}(P_{b})$ que en la
literatura se conoce como probabilidad crítica de Hammersley \citep{hammersley1957percolation}.
La variación de $P_{b}$ en el rango de $0$ a $1$determina la curva
de fase $P_{s,c}(P_{b})$ de la red $G$. 

El problema de percolación mixta sitio-enlace ha recibido considerable
atención en varias áreas y, tiene muchas aplicaciones en diferentes
campos. Por ejemplo, ha sido usado como modelo prototipo para el estudio
de la transición sol-gel (gelación) de polímeros \citep{Coniglio1979}.
En este modelo, los enlaces representan enlaces químicos, los sitios
ocupados representan monómeros, y los sitios no ocupados representan
moléculas solventes. También ha sido aplicado al estudio de la adsorción
disociativa sobre superficies metálicas \citep{Ziff1986,Rettner1987}.
En este sentido, Gao y Yang \citep{Gao1998} investigaron los procesos
de adsorción disociativa de dímeros y estudiaron las propiedades percolantes
de los monómeros disociados en función de la concentración de dímeros
(sitios) y la probabilidad de disociación (enlaces). Obtuvieron un
diagrama de fase que separa la región percolante de la no percolante.
Del mismo modo, la percolación mixta ha sido aplicada en investigaciones
biológicas y médicas. Un ejemplo típico aparece cuando alguna condición
patológica se expande por infección contagiosa a través de una gran
colonia estática de células (o individuos), de las cuales una proporción
$p$ es susceptible, y una proporción $1-p$ es inmune a la infección.
Las células están representadas por los vértices de una red (sitios)
que se conectan entre sí por los enlaces de la red, a través de los
cuales se puede propagar la infección \citep{Hammersley1980}. El
enfoque tradicional de la percolación sitio-enlace ha sido usado para
modelar la estructura de la red de poros y fenómenos de capilaridad.
En este marco, los sitios representan cuerpos porosos (cavidades)
y los enlaces corresponden a los cuellos de los poros (ventanas) \citep{Yanuka1990a,Englman1990}. 

El problema de percolación mixta sitio-enlace fue primero mencionado
por Frisch y Hammersley \citep{Frisch1963}, pero luego investigado
por primera vez por medio de simulaciones de Monte Carlo por Hoshen
et al \citep{hoshen1979percolation}. Agrawal et al \citep{Agrawal1979}
probaron que los exponentes críticos de la percolación pura de sitios,
son también válidos para la percolación mixta sitio-enlace. Nakanishi
y Reynolds \citep{Nakanishi1979} confirmaron esta conclusión a través
del estudio del grupo de renormalización del espacio de posiciones,
encontrando también que la frontera entre la zona de percolación y
no percolación en el plano $(P_{b},P_{s})$ esta gobernada por un
único punto fijo. Hammersley \citep{Hammersley1980} probó un teorema
para grafos parcialmente dirigidos, desde el cual se deriva la siguiente
desigualdad

\begin{equation}
P(P_{s}P_{b},1)\leq P(P_{s},P_{b})\leq P(1,P_{s}P_{b})\label{eq:desigualdad de hamersley}
\end{equation}

donde $P(P_{s},P_{b})$ es la probabilidad de percolación, es decir,
la probabilidad de que un único sitio fuente esté conectado a un conjunto
infinito de otros sitios. Esta desigualdad da cotas razonablemente
precisas para la probabilidad de percolación mixta. Para una red de
Bette Tarasevich et al \citep{Tarasevich1999} obtuvieron la relación 

\begin{equation}
P_{s}P_{b}=P_{c}\label{eq:curva de fase de bette}
\end{equation}

donde $P_{c}=P_{s,c}=P_{b,c}$.

La forma de la frontera entre la región percolante y no percolante,
fue subsecuentemente estudiada por Yanuka y Englman \citep{Yanuka1990}.
Ellos propusieron la siguiente ecuación para la curva crítica en el
plano $(P_{b},P_{s})$,

\begin{equation}
\frac{logP_{s}}{logP_{s,c}}+\frac{logP_{b}}{logP_{b,c}}=1\label{eq:ecuacion de yanuka-engelman}
\end{equation}

Esta ecuación satisface la desigualdad \ref{eq:desigualdad de hamersley}
y reproduce la Ecuación \ref{eq:curva de fase de bette} para la curva
de fase de la red de Bette. Esta ecuación define una curva en el plano
$(P_{b},P_{s})$, empezando en el punto $(P_{b,c},1)$ y finalizando
en $(1,P_{s,c})$. Yanuka y Englman \citep{Yanuka1990} mostraron
gráficas de la Ecuación \ref{eq:ecuacion de yanuka-engelman} además
de resultados de simulación para la curva de fase $(P_{b},P_{s})$
de las redes cuadrada, triangular, cúbica simple, y cúbica de cara
centrada (fcc). Todos los resultados de sus simulaciones se parecen
notablemente a la curva definida por esta ecuación. No es claro desde
su publicación, si los datos de por ejemplo, la red fcc, los cuales
yacen visiblemente por debajo de la curva crítica dada por la Ecuación
\ref{eq:ecuacion de yanuka-engelman}, son lo suficientemente precisos
para concluir que hay discrepancias entre ambas curvas.

Tretyakov e Inui \citep{tretyakov1995critical}, estudiaron la misma
curva crítica, pero ahora para percolación dirigida. Los resultados
de sus simulaciones fueron lo suficientemente precisos para concluir
que hay desviaciones del orden $10^{-3}$ a $10^{-4}$. Esta cuestión
despertó en Tarasevich et al \citep{Tarasevich1999}, el cuestionamiento
de si tales desviaciones también ocurren en la percolación sitio-enlace
regular. Por ello emprendieron un estudio sistemático de los valores
de umbral $(P_{b},P_{s})$ para percolación aleatoria sitio-enlace
en varias redes, para el modelo $S\cap B$. Presentaron resultados
de simulación los cuales se apartan mucho desde los valores obtenidos
por la Ecuación \ref{eq:ecuacion de yanuka-engelman} de Yanuka-Englman.
Las desviaciones más grandes se producen para redes donde el cociente
$\alpha=P_{s,c}/P_{b,c}$ es pequeño, es decir, en redes para las
cuales el umbral de percolación puro de enlace, es mucho mayor que
el umbral de percolación puro de sitios. Nosotros en este trabajo
usaremos en parte los resultados de la simulación de Tarasevich et
al \citep{Tarasevich1999}. En especial usaremos los datos que involucran
a las redes Arquimedianas $(4^{4}),(3^{6}),(6^{3}),(3^{3}.4^{2}),(3^{3}.4.3.4)$
y $(3.6.3.6)$, para validar nuestras simulaciones computacionales
en el caso del modelo $S\cap B$. También Tarasevich et al \citep{Tarasevich1999},
presentaron una ecuación aproximada para la curva de fase $(P_{b},P_{s})$
en el modelo $S\cap B$.

González et al \citep{gonzalez2013site} realizaron un estudio teórico
y computacional de la percolación mixta sitio-enlace sobre redes triangulares,
para ambos modelos, $S\cup B$ y $S\cap B$. Ellos obtuvieron la curva
de fase y los exponentes críticos de la transición de fase que ocurre
en estos sistemas. Además presentaron un enfoque teórico del problema
de percolación mixta sitio-enlace, basado en el cálculo exacto del
número de configuraciones sobre celdas triangulares finitas. En particular,
los puntos críticos $(P_{b},P_{s,c})$ de la curva de fase, fueron
obtenidos por el método propuesto por Yonezawa et al \citep{Yonezawa1989},
lo que hace este conjunto de datos de interés para la validación de
nuestras simulaciones en ambos modelos, $S\cap B$ y $S\cup B$. Por
una comunicación privada con los autores de este trabajo, también
contamos con la curva de fase asociada a las redes cuadradas, cuyos
puntos críticos fueron también obtenidos por el método propuesto por
Yonezawa et al \citep{Yonezawa1989}. Usamos tales datos para validar
los resultados de nuestras simulaciones para redes cuadradas para
ambos modelos, $S\cup B$ y $S\cap B$.

Una vez definido el modelo de percolación mixta, surge el problema
de cómo tratar geométricamente una red dada, de tal manera que sea
posible detectar cuando la misma ha percolado bajo el criterio de
percolación elegido, y que esta detección implique el menor esfuerzo
computacional posible. En respuesta a esta dificultad, veremos en
la Sección \ref{sec:Representaci=0000F3n-tradicional-de-las-RA},
cómo se trata tradicionalmente la red cuando se quiere detectar si
la misma ha percolado en alguna dirección predefinida, y en la Sección
\ref{sec:Representaci=0000F3n-alternativa-de-las-RA}, cómo se lo
puede hacer en \textit{forma alternativa}.

\section{\label{sec:Representaci=0000F3n-tradicional-de-las-RA}Representación
tradicional de las redes Arquimedianas}

\subsection{Origen y propiedades de las redes Arquimediana}

Sea $F=\left\{ T_{1},T_{2},...\right\} $ un teselado plano, es decir,
una familia contable de conjuntos cerrados o teselas $T_{i}$, los
cuales cubren el plano Euclidiano sin brechas o espacios ni traslapamientos
entre ellos. Mas explícitamente, la unión de los conjuntos o teselas
$T_{1},T_{2},...$ da como resultado el plano Euclidiano total, y
la intersección entre los interiores de los conjuntos $T_{1},T_{2},...,$
da como resultado el conjunto vacío \citep{grunbaum1987tilings}.

\begin{figure}
\begin{minipage}[t]{0.25\columnwidth}%
\begin{center}
\includegraphics[height=4cm]{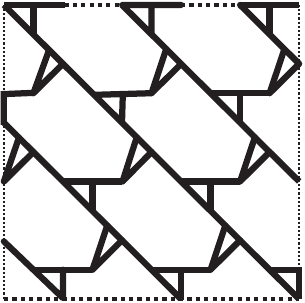}
\par\end{center}
\begin{center}
$(3.12^{2})$
\par\end{center}%
\end{minipage}%
\begin{minipage}[t]{0.25\columnwidth}%
\begin{center}
\includegraphics[height=4cm]{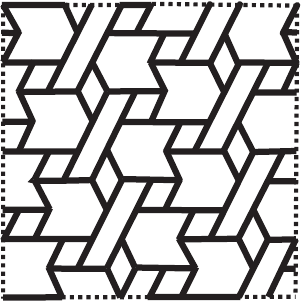}
\par\end{center}
\begin{center}
$(4.8^{2})$
\par\end{center}%
\end{minipage}%
\begin{minipage}[t]{0.25\columnwidth}%
\begin{center}
\includegraphics[height=4cm]{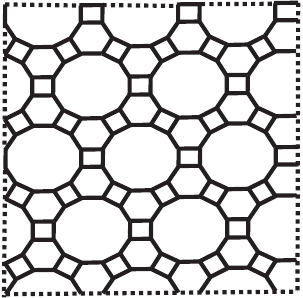}
\par\end{center}
\begin{center}
$(4.6.12)$
\par\end{center}%
\end{minipage}%
\begin{minipage}[t]{0.25\columnwidth}%
\begin{center}
\includegraphics[height=4cm]{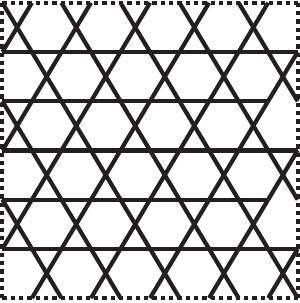}
\par\end{center}
\begin{center}
$(3,6.3.6)$
\par\end{center}%
\end{minipage}

\caption{\label{fig:Representaci=0000F3n-tradicional-de-Redes-Arquimedianas1}Representación
tradicional de las redes Arquimedianas. Los segmentos representan
enlaces, mientras las intersecciones de estos representan los sitios
de la red.}
\end{figure}

\begin{figure}
\begin{minipage}[t]{0.25\columnwidth}%
\begin{center}
\includegraphics[height=4cm]{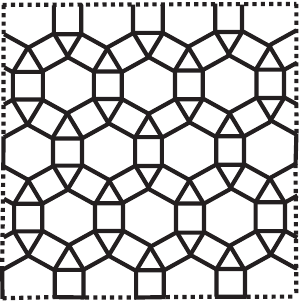}
\par\end{center}
\begin{center}
$(3.4.6.4)$
\par\end{center}%
\end{minipage}%
\begin{minipage}[t]{0.25\columnwidth}%
\begin{center}
\includegraphics[height=4cm]{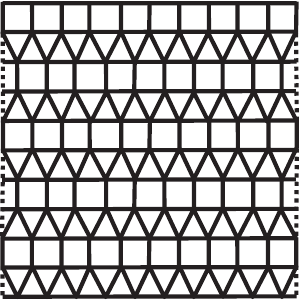}
\par\end{center}
\begin{center}
$(3^{3}.4^{2})$
\par\end{center}%
\end{minipage}%
\begin{minipage}[t]{0.25\columnwidth}%
\begin{center}
\includegraphics[height=4cm]{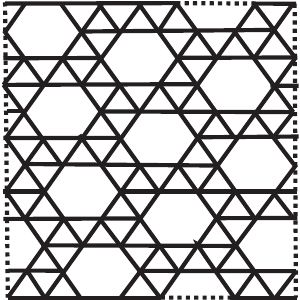}
\par\end{center}
\begin{center}
$(3^{4}.6)$
\par\end{center}%
\end{minipage}%
\begin{minipage}[t]{0.25\columnwidth}%
\begin{center}
\includegraphics[height=4cm]{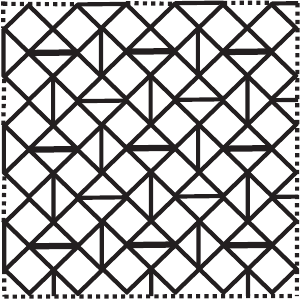}
\par\end{center}
\begin{center}
$(3^{2}.4.3.4)$
\par\end{center}%
\end{minipage}

\caption{\label{fig:Representaci=0000F3n-tradicional-de-Redes-Arquimedianas2}Representación
tradicional de las redes Arquimedianas. Los segmentos representan
enlaces, mientras las intersecciones de estos representan los sitios
de la red.}
\end{figure}

\begin{figure}
\begin{centering}
\begin{minipage}[t]{0.3\columnwidth}%
\begin{center}
\includegraphics[height=4cm]{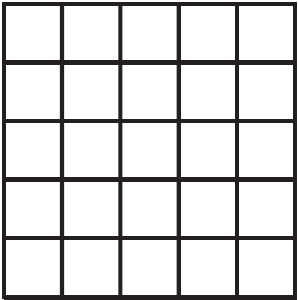}
\par\end{center}
\begin{center}
$(4^{4})$
\par\end{center}%
\end{minipage}%
\begin{minipage}[t]{0.3\columnwidth}%
\begin{center}
\includegraphics[height=4cm]{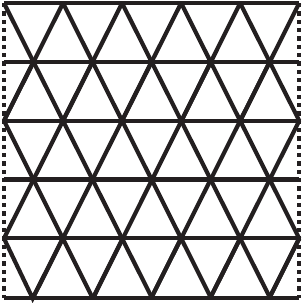}
\par\end{center}
\begin{center}
$(3^{6})$
\par\end{center}%
\end{minipage}%
\begin{minipage}[t]{0.3\columnwidth}%
\begin{center}
\includegraphics[height=4cm]{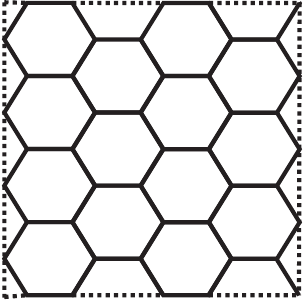}
\par\end{center}
\begin{center}
$(6^{3})$
\par\end{center}%
\end{minipage}
\par\end{centering}
\caption{\label{fig:Representaci=0000F3n-tradicional-de-Redes-Arquimedianas3}Representación
tradicional de las redes Arquimedianas. Los segmentos representan
enlaces, mientras las intersecciones de estos representan los sitios
de la red.}
\end{figure}

Si requerimos que el teselado plano $F$, solo este compuesto de polígonos
regulares, y que todos sus vértices sean del mismo tipo, entonces,
existen precisamente $11$ teselados distintos que cumplen este requerimiento
\citep{grunbaum1987tilings}. Estos teselados son $(3^{6})$, $(3^{4}.6)$,
$(3^{2}.4^{2})$, $(3^{2}.4.3.4)$, $(3.4.6.4)$, $(3.6.3.6)$, $(3.12^{2})$,
$(4^{4})$, $(4.6.12)$, $(4.8^{2})$ y $(6^{3})$, notación o símbolo
que indica el único tipo de vértice que los compone \citep{grunbaum1987tilings}.
En las Figuras \ref{fig:Representaci=0000F3n-tradicional-de-Redes-Arquimedianas1},
\ref{fig:Representaci=0000F3n-tradicional-de-Redes-Arquimedianas2},
y \ref{fig:Representaci=0000F3n-tradicional-de-Redes-Arquimedianas3},
se muestran los $11$ teselados listados aquí. Usualmente son llamados
Teselados Arquimedianos o Redes Arquimedianas, aunque algunos autores
los llaman Teselados Homogéneos o Semiregulares. El adjetivo ''Arquimediano''
se refiere a que estos teselados son monogonales, es decir, los vecinos
inmediatos de cualquier par de vértices ''ven lo mismo''. También
estos teselados tienen la propiedad de ser isogonales, lo que significa
que cada par de vértices son equivalentes bajo alguna simetría del
teselado \citep{grunbaum1987tilings}.

Al ser cada una de las redes Arquimedianas construidas con un solo
tipo de vértice, todos los vértices pertenecientes a una red dada,
tienen el mismo grado o número de coordinación z, supuesto una red
de tamaño infinito, o bien finita con condiciones de contorno periódicas
(CCP). 

Se caracterizan además por ser planas \citep{godsil2001algebraic},
es decir, todos los puntos de intersección de sus enlaces constituyen
sus vértices, los cuales podrán ocuparse o no con elementos físicos
de interés. También son planas en el sentido de la teoría de grafos
\citep{godsil2001algebraic}, pues se pueden dibujar en el plano sin
que ningún enlace o arista se cruce con otro. Todas ellas pueden ser
dibujadas sobre esferas, siendo esto una propiedad de los grafos planos
en general.

Hacemos notar aquí que, elegido un vértice de una red Arquimediana
dada, y un sentido de giro que sea el mismo para todos sus vértices,
todos los vértices generarán el mismo nombre o notación para la red
elegida, excepto para la red conocida como $(4.6.12)$. Si observamos
la red (4.6.12) en la Figura \ref{fig:Representaci=0000F3n-tradicional-de-Redes-Arquimedianas1},
elegimos un sentido de giro, y luego recorremos cada uno de los vértices
de un dodecágano cualquiera, no todos sus vértices generan literalmente
el mismo nombre o notación, aunque una vez ordenados estos nombres
lexicográficamente, si lo hacen . 

\subsection{Celdas de construcción tradicionales de las redes Arquimedianas}

\begin{figure}
\begin{centering}
\begin{minipage}[t]{0.25\columnwidth}%
\begin{center}
\includegraphics[width=35mm]{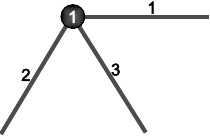}
\par\end{center}
\begin{center}
CU$(3^{6})$
\par\end{center}%
\end{minipage}%
\begin{minipage}[t]{0.25\columnwidth}%
\begin{center}
\includegraphics[width=35mm]{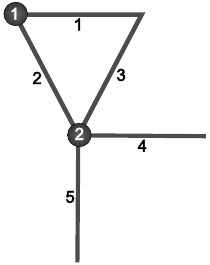}
\par\end{center}
\begin{center}
CU(3$^{3}$.4$^{2}$)
\par\end{center}%
\end{minipage}%
\begin{minipage}[t]{0.25\columnwidth}%
\begin{center}
\includegraphics[width=35mm]{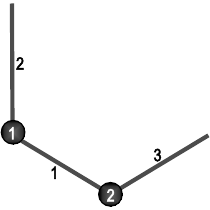}
\par\end{center}
\begin{center}
CU$(6^{3})$
\par\end{center}%
\end{minipage}%
\begin{minipage}[t]{0.25\columnwidth}%
\begin{center}
\includegraphics[width=35mm]{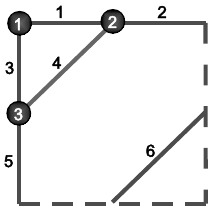}
\par\end{center}
\begin{center}
CU$(3.6.3.6)$
\par\end{center}%
\end{minipage}
\par\end{centering}
\caption{\label{fig:CU-tradi-1}Cuatro de las 11 celdas unidad (CU) de las
redes Arquimedianas. Cada CU debe ser trasladada en dos direcciones
convenientemente elegidas, para obtener la red deseada.}
\end{figure}

\begin{figure}
\begin{centering}
\begin{minipage}[t]{0.25\columnwidth}%
\begin{center}
\includegraphics[width=35mm]{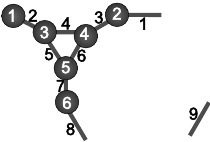}
\par\end{center}
\begin{center}
CU$(3.12^{2})$
\par\end{center}%
\end{minipage}%
\begin{minipage}[t]{0.25\columnwidth}%
\begin{center}
\includegraphics[width=35mm]{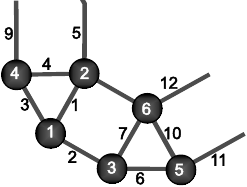}
\par\end{center}
\begin{center}
CU$(3.4.6.4)$
\par\end{center}%
\end{minipage}%
\begin{minipage}[t]{0.25\columnwidth}%
\begin{center}
\includegraphics[width=35mm]{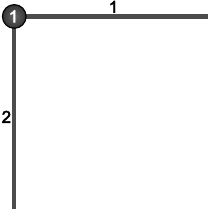}
\par\end{center}
\begin{center}
CU(4$^{4}$)
\par\end{center}%
\end{minipage}%
\begin{minipage}[t]{0.25\columnwidth}%
\begin{center}
\includegraphics[width=35mm]{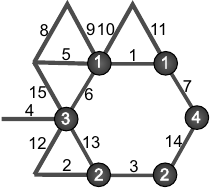}
\par\end{center}
\begin{center}
CU$(3^{4}.6)$
\par\end{center}%
\end{minipage}
\par\end{centering}
\caption{\label{fig:CU-tradi-2}Cuatro de las 11 celdas unidad (CU) de las
redes Arquimedianas. Cada CU debe ser trasladada en dos direcciones
convenientemente elegidas, para obtener la red deseada.}
\end{figure}

\begin{figure}
\begin{centering}
\begin{minipage}[t]{0.3\columnwidth}%
\begin{center}
\includegraphics[width=35mm]{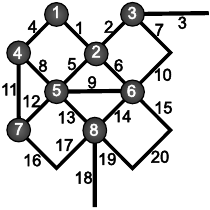}
\par\end{center}
\begin{center}
CU$(3^{2}.4.3.4)$
\par\end{center}%
\end{minipage}%
\begin{minipage}[t]{0.3\columnwidth}%
\begin{center}
\includegraphics[width=35mm]{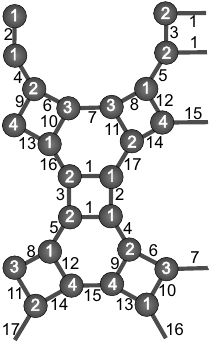}
\par\end{center}
\begin{center}
CU(4.6.12)
\par\end{center}%
\end{minipage}%
\begin{minipage}[t]{0.3\columnwidth}%
\begin{center}
\includegraphics[width=35mm]{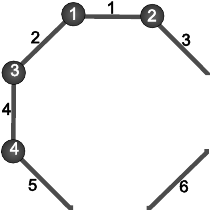}
\par\end{center}
\begin{center}
CU$(4.8^{2})$
\par\end{center}%
\end{minipage}
\par\end{centering}
\caption{\label{fig:CU-tradi-3}Tres de las 11 celdas unidad (CU) de las redes
Arquimedianas. Cada CU debe ser trasladada en dos direcciones convenientemente
elegidas, para obtener la red deseada.}
\end{figure}

Es habitual que la construcción geométrica de las redes Arquimedianas,
se realice a partir de celdas de construcción llamadas celdas unidad
(CU). En las Figuras \ref{fig:CU-tradi-1}, \ref{fig:CU-tradi-2},
y \ref{fig:CU-tradi-3}, se han representado las CU de cada una de
ellas. Los sitios o vértices se han representado con círculos llenos
de color gris oscuro y etiqueta numérica de color blanco, y los enlaces
se han representado con segmentos de color gris y etiqueta numérica
de color negro. Una red Arquimediana cualquiera, se obtiene trasladando
su respectiva CU en dos direcciones distintas, no necesariamente perpendiculares
entre si, tantas veces como sea necesario hasta obtener el tamaño
deseado de las aristas de la red. 

Que dos sitios tengan la misma numeración en una dada CU, significa
que sus primeros y segundos vecinos pueden ser identificados con los
mismos vectores desplazamientos, respecto de la posición de cada sitio
de igual numeración, supuesto una red de tamaño infinito, o de tamaño
finito con condiciones de contorno periódicas (CCP). En este sentido
decimos que los sitios de igual numeración son equivalentes entre
sí. Igualmente, que dos enlaces de una dada CU tengan la misma numeración,
significa que son equivalentes en el mismo sentido explicado para
los sitios de igual numeración. Como veremos, esta equivalencia es
clara en el contexto de la \textit{representación alternativa} de
una red, la cual explicamos en una sección posterior, pero no así
en el contexto de la representación tradicional. 

\subsection{Tratamiento geométrico tradicional de una red}

Ambos modelos $(S\cap B$ y $S\cup B$) son convertidos a redes efectivas
de enlaces \citep{gonzalez2013site,gonzalez2016site}, convirtiendo
el problema de percolación mixta, en un problema de percolación puro
de enlaces, donde la operación computacional preponderante es el análisis
de conexión entre un enlace dado, y los enlaces primeros vecinos de
este. A continuación describimos cómo se hace la transformación de
la red real a una red efectiva.

\subsubsection{Red efectiva en el modelo S$\cap$B}

\begin{figure}
\begin{centering}
\includegraphics[width=8cm]{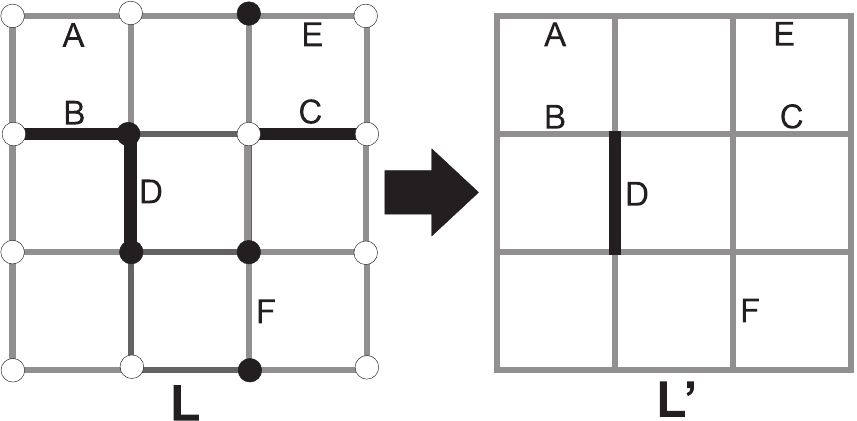}
\par\end{centering}
\caption{\label{fig:Red-efectiva-mod-S-AND-B}Transformación de la red real
L en la red efectiva L' en el modelo $S\cap B$. Los círculos llenos
de color negro indican sitios ocupados. Los enlaces representados
con líneas en negro indican que están ocupados. Los círculos no llenos
indican sitios vacíos. Las líneas grises indican enlaces desocupados.}
\end{figure}

En este modelo la red de sitios y enlaces, L, puede puede ser convertida
en una red efectiva de solamente enlaces, L', ver Figura \ref{fig:Red-efectiva-mod-S-AND-B}.
El mapeo de L en L' sigue el siguiente proceso:
\begin{enumerate}
\item Cada enlace vacío de L se transforma en un enlace vacío de L'. Ver
etiquetas A, E, y F de la Figura \ref{fig:Red-efectiva-mod-S-AND-B}.
\item Cada enlace ocupado con uno o dos sitios extremos vacíos de L, se
transforma en un enlace vacío de L'. Ver etiquetas B y C de la Figura
\ref{fig:Red-efectiva-mod-S-AND-B}
\item Cada enlace ocupado con sus sitios extremos ocupados de L, se transforma
en un enlace ocupado de L'. Ver etiqueta D de la Figura \ref{fig:Red-efectiva-mod-S-AND-B}
\end{enumerate}
Por lo tanto, un problema de percolación de sitios y enlaces L, se
transforma en un problema de percolación de solo enlaces en la red
L'.

\subsubsection{Red efectiva en el modelo S$\cup$B}

\begin{figure}
\centering{}\includegraphics[width=8cm]{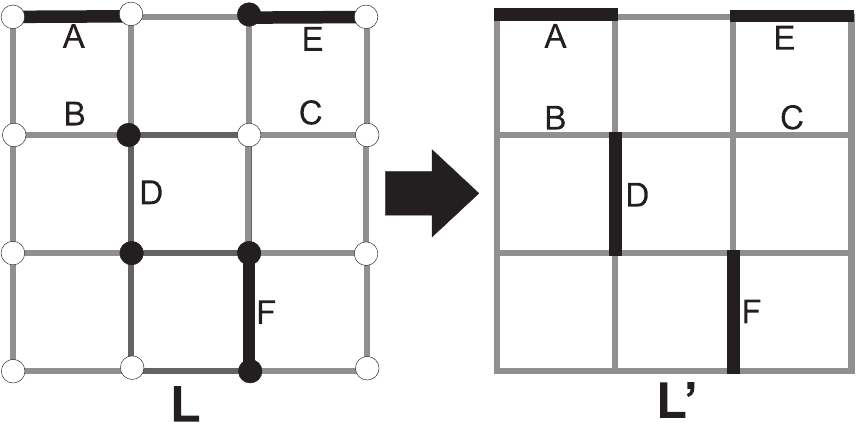}\caption{\label{fig:Red-efectiva-mod-S-OR-B}Transformación de la red real
L en la red efectiva L' en el modelo $S\cup B$. Los círculos llenos
de color negro indican sitios ocupados. Los enlaces representados
con líneas negras indican que están ocupados. Los círculos no llenos
indican sitios vacíos. Las líneas grises indican enlaces desocupados.}
\end{figure}

En este modelo la red de sitios y enlaces, L, puede ser convertida
en una red efectiva de solamente enlaces, L', ver Figura \ref{fig:Red-efectiva-mod-S-OR-B}.
Ahora, el mapeo de L en L' sigue el siguiente proceso:
\begin{enumerate}
\item Cada enlace ocupado de L se transforma en un enlace ocupado de L'.
Ver etiquetas A, E, y F de la Figura \ref{fig:Red-efectiva-mod-S-OR-B}.
\item Cada enlace vacío con uno o dos sitios extremos vacíos de L, se transforma
en un enlace vacío de L'. Ver etiquetas B y C de la Figura \ref{fig:Red-efectiva-mod-S-OR-B}.
\item Cada enlace vacío con sus sitios extremos ocupados de L, se transforma
en un enlace ocupado de L'. Ver etiqueta D de la Figura \ref{fig:Red-efectiva-mod-S-OR-B}.
\end{enumerate}
Nuevamente, un problema de percolación de sitios y enlaces L, se transforma
en un problema de percolación de solo enlaces en la red L'.

En cualquiera de los modelos, una vez que se ha completado el mapeo,
cada configuración percolante o no percolante en la red efectiva,
$L'$, corresponderá a una configuración percolante o no percolante
en la red original, $L$. También, la red efectiva, $L'$, implica
una interacción a primeros vecinos entre enlaces. 

Para la realización computacional de un análisis de percolación en
la red efectiva $L'$, hace falta un método sistemático de etiquetado
de los enlaces que componen la misma. En la literatura solo se vislumbran
indexaciones ad hoc, cada una de ellas válida para la red particular
que se creo. 

Sigue al etiquetado de los enlaces en la red efectiva $L'$, la aplicación
de algún algoritmo de análisis de clústeres y de detección de percolación
como el de Hoshen y Kopelman \citep{hoshen1979percolation}, o bien
alguna variante de los algoritmos unión/encontrar que se encuentran
en la literatura de la física computacional, o bien en la literatura
de las ciencias de la computación \citep{sedgewick1995algoritmos}.

\section{\label{sec:Representaci=0000F3n-alternativa-de-las-RA}Representación
alternativa de las redes Arquimedianas}

\subsection{\label{subsec:CU-alternativas}Celdas de construcción alternativas
de las redes Arquimedianas}

\begin{figure}
\centering{}%
\begin{minipage}[t]{0.15\columnwidth}%
\begin{center}
\includegraphics{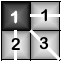}
\par\end{center}
\begin{center}
$(3^{6})$
\par\end{center}%
\end{minipage}%
\begin{minipage}[t]{0.15\columnwidth}%
\begin{center}
\includegraphics{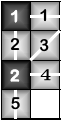}
\par\end{center}
\begin{center}
$(3^{3}.4^{2})$
\par\end{center}%
\end{minipage}%
\begin{minipage}[t]{0.25\columnwidth}%
\begin{center}
\includegraphics{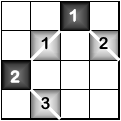}
\par\end{center}
\begin{center}
$(6^{3})$
\par\end{center}%
\end{minipage}%
\begin{minipage}[t]{0.25\columnwidth}%
\begin{center}
\includegraphics{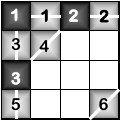}
\par\end{center}
\begin{center}
$(3.6.3.6)$
\par\end{center}%
\end{minipage}%
\begin{minipage}[t]{0.15\columnwidth}%
\begin{center}
\includegraphics{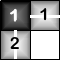}
\par\end{center}
\begin{center}
$(4^{4})$
\par\end{center}%
\end{minipage}\caption{\label{fig:CU-Alter-1}Cinco de las 11 CU de las redes Arquimedianas.
Todas las redes Arquimedianas se construyen trasladando horizontalmente
y verticalmente las respectivas CU.}
\end{figure}

\begin{figure}
\begin{minipage}[t]{0.25\columnwidth}%
\begin{center}
\includegraphics{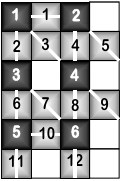}
\par\end{center}
\begin{center}
$(3.4.6.4)$
\par\end{center}%
\end{minipage}%
\begin{minipage}[t]{0.5\columnwidth}%
\begin{center}
\includegraphics{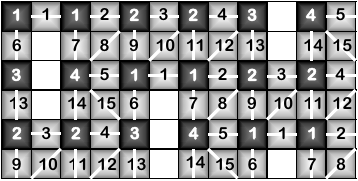}
\par\end{center}
\begin{center}
$(3^{4}.6)$
\par\end{center}%
\end{minipage}%
\begin{minipage}[t]{0.25\columnwidth}%
\begin{center}
\includegraphics[scale=0.9]{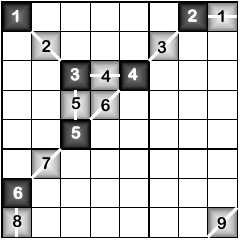}
\par\end{center}
\begin{center}
$(3.12^{2})$
\par\end{center}%
\end{minipage}

\caption{\label{fig:CU-Alter-2}Tres de las 11 CU de las redes Arquimedianas.
Todas las redes Arquimedianas se construyen trasladando horizontalmente
y verticalmente las respectivas CU.}
\end{figure}

\begin{figure}
\begin{centering}
\begin{minipage}[t]{0.25\columnwidth}%
\begin{center}
\includegraphics{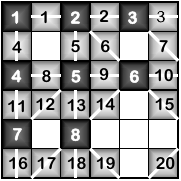}
\par\end{center}
\begin{center}
$(3^{2}.4.3.4)$
\par\end{center}%
\end{minipage}%
\begin{minipage}[t]{0.5\columnwidth}%
\begin{center}
\includegraphics[angle=90]{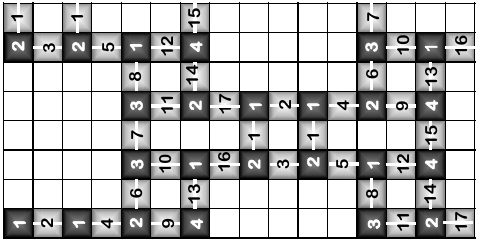}
\par\end{center}
\begin{center}
$(4.6.12)$
\par\end{center}%
\end{minipage}%
\begin{minipage}[t]{0.25\columnwidth}%
\begin{center}
\includegraphics{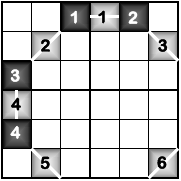}
\par\end{center}
\begin{center}
$(4.8^{2})$
\par\end{center}%
\end{minipage}
\par\end{centering}
\caption{\label{fig:CU-Alter-3}Tres de las 11 CU de las redes Arquimedianas.
Todas las redes Arquimedianas se construyen trasladando horizontalmente
y verticalmente las respectivas CU.}
\end{figure}

\begin{figure}
\raggedright{}%
\begin{minipage}[t]{0.5\columnwidth}%
\begin{center}
\includegraphics[width=8cm]{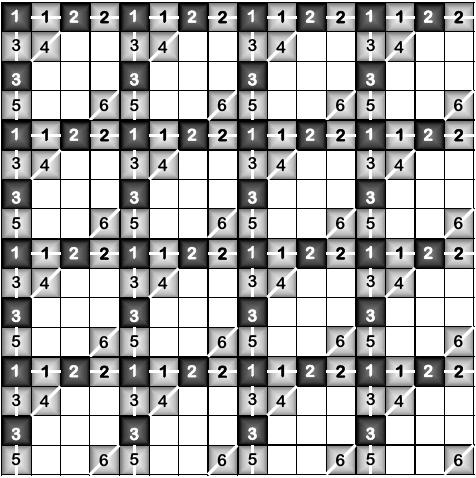}
\par\end{center}
\begin{center}
\textbf{(a)} $(3.6.3.6)$ de 4CU x 4CU
\par\end{center}%
\end{minipage}%
\begin{minipage}[t]{0.5\columnwidth}%
\begin{center}
\includegraphics[width=8cm]{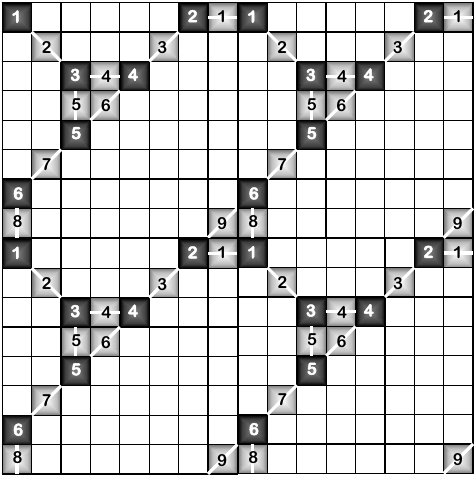}
\par\end{center}
\begin{center}
\textbf{(b)} $(3.12^{2})$ de 2CU x 2CU
\par\end{center}%
\end{minipage}\caption{\label{fig:Ejemplo-de-redes-alternativas}Ejemplo de redes alternativas.
\textbf{(a)} kagome o $(3.6.3.6)$ de $4CU\times4CU$ y \textbf{(b)}
K+ o $(3.12^{2})$ de $2CU\times2CU$.}
\end{figure}

En las Figuras \ref{fig:CU-Alter-1}, \ref{fig:CU-Alter-2}, y \ref{fig:CU-Alter-3},
mostramos las CU de cada una de las redes Arquimedianas, pero ahora
representadas de la manera propuesta en este trabajo. Tanto los sitios
como los enlaces se representan mediante cuadrados de arista \textquotedbl unidad\textquotedbl .
La representación de sitios de una red mediante cuadrados no es nueva,
siendo estos cuadrados parte de un cuadriculado más grande el cual
representa la red que queremos estudiar. Por ejemplo, Kim Christensen
y Nicholas R. Moloney\citep{christensen2005complexity}, usaron esta
representación para ilustrar ciertos aspectos del problema de percolación
pura de sitios en una red unidimensional y en la red cuadrada, pero
no profundizaron sobre esta representación. También es habitual encontrar
esta representación de la red cuadrada mediante un cuadriculado, cuando
se quiere ilustrar un proceso de percolación pura de sitios sobre
la pantalla de un monitor de PC o de una notebook. No obstante, nuestra
propuesta tiene algunos matices. Observemos que:
\begin{enumerate}
\item Tanto los sitios como los enlaces están representados por celdas cuadradas
de arista \textquotedbl unidad\textquotedbl . Los sitios están pintados
con un degradado que va de negro a gris (50\%), mientras que los enlaces
están pintados con un degradado que va de gris (50\%) a blanco. El
resto de las celdas representan \textquotedbl huecos\textquotedbl{}
no usados por la red en estudio. 
\item Todas las CU son cuadradas o rectangulares.
\item Los enlaces solamente enlazan sitios en las direcciones, horizontal,
vertical, en diagonal a $45^{o}$ o en diagonal a $-45^{o}$. La dirección
en la cual actúa un enlace se señala con una línea blanca. Dada una
CU, un enlace solo puede unir un único par de sitios, es decir, opera
en una única dirección. 
\item Como en el modelo tradicional, un sitio podrá tener como primer vecino
uno o varios enlaces, mientras que un enlace podrá tener como primer
vecino hasta dos sitios a lo sumo. 
\item A diferencia del modelo tradicional, los sitios y los enlaces no pierden
su identidad como tal al tratar la red, es decir, los sitios siguen
siendo sitios y los enlaces siguen siendo enlaces. Lo que se ha cambiado
es su representación geométrica.
\item Todas las redes se construyen trasladando la CU correspondiente, en
dirección horizontal o vertical, tantas veces como sea necesario,
hasta alcanzar el tamaño de aristas deseado. A modo de ejemplo, en
la Figura\  \ref{fig:Ejemplo-de-redes-alternativas}\textbf{(a)},
se muestra la red de kagome o $(3.6.3.6)$ de 4CU x 4CU, y en la Figura
\  \ref{fig:Ejemplo-de-redes-alternativas}\textbf{(b)}, se muestra
la red K+ o $(3.12^{2})$ de 2CU x 2CU.
\item En esta representación podemos hablar de la densidad absoluta de sitios
(enlaces) o de celdas posibles de ser ocupados por sitios (enlaces),
la cual queda claramente determinada por la estructura de la CU, la
que a su vez también determina la densidad absoluta de huecos de la
red. La traslación de la CU no afecta la densidad de sitios-enlaces-huecos
de la red.
\item En esta representación es posible construir cuadriculados de tamaño
mínimo $\mathbf{L_{min}\times L_{min}=48\times48}$ dentro de los
cuales podemos encajar un número entero de veces cada una de las CU
de las redes Arquimedianas. Donde $\mathbf{L_{min}=48}$ es el MCM
(Mínimo Común Múltiplo) de todos los tamaños de aristas implicados
en las respectivas CU, teniendo en cuenta filas y columnas de las
mismas. Es decir, $\mathbf{L_{min}=MCM\left(2,4,6,8,12,16\right)=2^{4}\times3=48}$.
El hecho de que se pueda estandarizar el tamaño de las redes Arquimedianas
utilizando múltiplos de 48, facilita en gran medida el estudio y la
comparación de estas redes en el contexto de \textit{problemas de
competencias o competición}. La estandarización de las dimensiones
de las redes permite una comparación más justa y objetiva entre ellas,
ya que elimina las disparidades en términos de tamaño.
\item En esta representación se vislumbra un nuevo tipo de problema de percolación,
el cual puede en principio tener aplicaciones prácticas: \\
Consideremos un cuadriculado de tamaño $\mathbf{L\times L}$, en el
cual se inscribe por ejemplo la red $(3.6.3.6)$, donde $\mathbf{L}$
es un múltiplo de 48. Inicialmente todos los cuadrados de la red están
\textquotedbl desactivados\textquotedbl . Ahora procedamos a ''activar''
cada cuadrado aleatoriamente, hasta alcanzar cierta concentración
deseada $P_{a}$ de cuadrados activos. Si el cuadrado activado corresponde
a un sitio o a un enlace de la red $(3.6.3.6)$ se ocupa , y si corresponde
a un hueco no se hace nada. ¿Para qué concentración crítica $P_{a,c}$
de cuadrados activos, esta red \textquotedbl percola\textquotedbl{}
en alguno de los modelos $S\cup B$ o $S\cap B$?. En el límite termodinámico
$L\rightarrow\infty$ podremos obtener el valor crítico $P_{a,c}^{\infty}$.
Si contestamos esta pregunta para cada una de las redes Arquimedianas,
podremos comparar sus valores críticos $P_{a,c}$ o bien $P_{a,c}^{\infty}$
y sacar algunas conclusiones.
\item Otro problema: \\
Consideremos un cuadriculado de tamaño $\mathbf{L\times L}$, en el
cual se inscribe por ejemplo la red $(3.3.6.3)$, donde $\mathbf{L}$
es un múltiplo de 48. Inicialmente todos los cuadrados de la red están
''activados''. Ahora procedamos a \textquotedbl desactivar\textquotedbl{}
cada cuadrado aleatoriamente, hasta alcanzar cierta concentración
deseada $P_{a}$ de cuadrados desactivados. Si el cuadrado desactivado
corresponde a un sitio o a un enlace de la red $(3.6.3.6)$ se desocupa,
y si corresponde a un hueco no se hace nada. ¿Para qué concentración
crítica $P_{a,c}$ de cuadrados desactivados, la red deja de \textquotedbl percolar\textquotedbl{}
en alguno de los modelos $S\cup B$ o $S\cap B$?. En el límite termodinámico
$L\rightarrow\infty$ podremos obtener el valor crítico $P_{a,c}^{\infty}$.
Si contestamos esta pregunta para cada una de las redes Arquimedianas,
podremos comparar sus valores críticos $P_{a,c}$ \ o bien $P_{a,c}^{\infty}$,
y sacar algunas conclusiones. 
\end{enumerate}

\subsection{\label{subsec:Etiquetado-y-caracterizaci=0000F3n}Etiquetado y caracterización
de cada uno de los sitios y enlaces de una red Arquimediana}

\subsubsection{Etiquetado}

Resulta obvio en esta representación que cada sitio y enlace de la
red se puede etiquetar con un par ordenado de números enteros $\left(f,c\right)$,
donde f es la f-ésima fila donde se localiza la celda que representa
el sitio o enlace de la red, y c es la c-ésima columna sobre la red
considerada. Las filas se cuentan de arriba hacia abajo, y las columnas
de izquierda a derecha, tal como se hace con una matriz de números
en el contexto de las matemáticas.

\subsubsection{Caracterización}

Tomamos como ejemplo, la red de kagome (3,6,3,6) de la Figura \ref{fig:Ejemplo-de-redes-alternativas}\textbf{(a)}.
Llamamos $NFCU$ al número de filas de la celda unidad, y $NCCU$
al número de columnas de la misma. Los sitios enumerados con 1 en
la red verifican que $f_{i}\mod NFCU=0$, y $c_{i}\mod NCCU=0$; los
enumerados con 2 verifican que $f_{i}\mod NFCU=0$, y $c_{i}\mod NCCU=2$;
y los numerados con 3 verifican que $f_{i}\mod NFCU=2$, y $c_{i}\mod NCCU=0$.
De la misma forma, se pueden caracterizar los seis tipos de enlaces
que componen la CU. Los enlaces numerados con 1 verifican que $f_{i}\mod NFCU=0$,
y $c_{i}\mod NCCU=1$; los numerados con 2 verifican que $f_{i}\mod NFCU=0$,
y $c_{i}\mod NCCU=3$; los numerados con 3 verifican que $f_{i}\mod NFCU=1$,
y $c_{i}\mod NCCU=0$; los numerados con 4 verifican que $f_{i}\mod NFCU=1$,
y $c_{i}\mod NCCU=1$; los numerados con 5 verifican que $f_{i}\mod NFCU=0$,
y $c_{i}\mod NCCU=3$; y los numerados con 6 verifican que $f_{i}\mod NFCU=3$,
y $c_{i}\mod NCCU=3$. En la Tabla \ref{tab:Caracterizaci=0000F3n de los sitios y enlaces de la CU de la red (3.6.3.6)},
resumimos la caracterización de los sitios y enlaces de la CU de la
red de kagome. Por lo tanto, la función módulo resuelve el problema
de la distinción de los distintos tipos de sitios y enlaces que componen
la CU, y que por ende también componen la red al ser esta obtenida
por traslación repetida de la CU, tanto en dirección horizontal como
vertical. Entonces, el etiquetado, la estructura de la CU, en conjunción
con la función módulo, caracterizan el tipo de sitios y enlaces que
componen la red.

\begin{table}
\begin{centering}
\noindent\begin{minipage}[t]{1\columnwidth}%
\begin{center}
\includegraphics[scale=1.2]{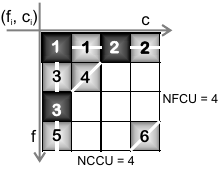}
\par\end{center}
\begin{center}
\begin{tabular}{|c|c|c|c|c|c|c|c|c|c|}
\hline 
CU$(3.6.3.6)$ & $s_{1}$ & $s_{2}$ & $s_{3}$ & $b_{1}$ & $b_{2}$ & $b_{3}$ & $b_{4}$ & $b_{5}$ & $b_{6}$\tabularnewline
\hline 
\hline 
$f_{i}\mod NFCU$ & $0$ & $0$ & $2$ & $0$ & $0$ & $1$ & $1$ & $3$ & $3$\tabularnewline
\hline 
$c_{i}\mod NCCU$ & $0$ & $2$ & $0$ & $1$ & $3$ & $0$ & $1$ & $0$ & $3$\tabularnewline
\hline 
\end{tabular}
\par\end{center}%
\end{minipage}
\par\end{centering}
\caption{\label{tab:Caracterizaci=0000F3n de los sitios y enlaces de la CU de la red (3.6.3.6)}Caracterización
de los sitios y enlaces de la CU de la red $(3.6.3.6)$ mediante la
función módulo.}
\end{table}

\subsubsection{Hoja de datos}

\begin{figure}
\begin{centering}
\begin{tabular}{|c||c||c||c|c||c||c||c|c||c||c||c|}
\hline 
\multicolumn{12}{|c|}{\textbf{(3.6.3.6)}}\tabularnewline
\hline 
\hline 
\multicolumn{4}{|c|}{\textbf{\scriptsize{}Tipo de Vértice}} & \multicolumn{4}{c|}{\textbf{\scriptsize{}CU Estándar}} & \multicolumn{4}{c|}{\textbf{\scriptsize{}CU Alternativo}}\tabularnewline
\hline 
\multicolumn{4}{|c|}{%
\begin{minipage}[t][30mm][c]{30mm}%
\begin{center}
\includegraphics{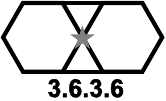}
\par\end{center}%
\end{minipage}} & \multicolumn{4}{c|}{%
\begin{minipage}[t][30mm][c]{30mm}%
\begin{center}
\includegraphics[scale=0.5]{CU-3C6C3C6-TRADIC-GRIS}
\par\end{center}%
\end{minipage}} & \multicolumn{4}{c|}{%
\begin{minipage}[t][30mm][c]{30mm}%
\begin{center}
\includegraphics{CU-3C6C3C6-ALTER-GRIS}
\par\end{center}%
\end{minipage}}\tabularnewline
\hline 
\multicolumn{12}{|c|}{}\tabularnewline
\hline 
\end{tabular}
\par\end{centering}
\smallskip{}

\centering{}%
\begin{tabular}{|c|c|c|c|c|c|c|c|c|c|c|c|}
\hline 
\multicolumn{12}{|c|}{\textbf{\scriptsize{}Sitios}}\tabularnewline
\hline 
\multicolumn{4}{|c|}{\textbf{\scriptsize{}Vecinos}} & \multicolumn{4}{c|}{\textbf{\scriptsize{}1NN}} & \multicolumn{4}{c|}{\textbf{\scriptsize{}2NN}}\tabularnewline
\hline 
{\scriptsize{}s} & {\scriptsize{}1} & {\scriptsize{}f} & {\scriptsize{}=} & {\scriptsize{}-1} & {\scriptsize{}0} & {\scriptsize{}+1} & {\scriptsize{}0} & {\scriptsize{}-2} & {\scriptsize{}0} & {\scriptsize{}+2} & {\scriptsize{}0}\tabularnewline
\hline 
{\scriptsize{}s} & {\scriptsize{}1} & {\scriptsize{}c} & {\scriptsize{}=} & {\scriptsize{}0} & {\scriptsize{}-1} & {\scriptsize{}0} & {\scriptsize{}+1} & {\scriptsize{}0} & {\scriptsize{}-2} & {\scriptsize{}0} & {\scriptsize{}+2}\tabularnewline
\hline 
{\scriptsize{}s} & {\scriptsize{}2} & {\scriptsize{}f} & {\scriptsize{}=} & {\scriptsize{}0} & {\scriptsize{}+1} & {\scriptsize{}0} & {\scriptsize{}-1} & {\scriptsize{}0} & {\scriptsize{}+2} & {\scriptsize{}0} & {\scriptsize{}-2}\tabularnewline
\hline 
{\scriptsize{}s} & {\scriptsize{}2} & {\scriptsize{}c} & {\scriptsize{}=} & {\scriptsize{}-1} & {\scriptsize{}-1} & {\scriptsize{}+1} & {\scriptsize{}+1} & {\scriptsize{}-2} & {\scriptsize{}-2} & {\scriptsize{}+2} & {\scriptsize{}+2}\tabularnewline
\hline 
{\scriptsize{}s} & {\scriptsize{}3} & {\scriptsize{}f} & {\scriptsize{}=} & {\scriptsize{}-1} & {\scriptsize{}+1} & {\scriptsize{}+1} & {\scriptsize{}-1} & {\scriptsize{}-2} & {\scriptsize{}+2} & {\scriptsize{}+2} & {\scriptsize{}-2}\tabularnewline
\hline 
{\scriptsize{}s} & {\scriptsize{}3} & {\scriptsize{}c} & {\scriptsize{}=} & {\scriptsize{}0} & {\scriptsize{}-1} & {\scriptsize{}0} & {\scriptsize{}+1} & {\scriptsize{}0} & {\scriptsize{}-2} & {\scriptsize{}0} & {\scriptsize{}+2}\tabularnewline
\hline 
\multicolumn{12}{|c|}{\textbf{\scriptsize{}Enlaces}}\tabularnewline
\hline 
\multicolumn{4}{|c|}{\textbf{\scriptsize{}Vecinos}} & \multicolumn{2}{c|}{\textbf{\scriptsize{}1NN}} & \multicolumn{6}{c|}{\textbf{\scriptsize{}2NN}}\tabularnewline
\hline 
{\scriptsize{}e} & {\scriptsize{}1} & {\scriptsize{}f} & {\scriptsize{}=} & {\scriptsize{}0} & {\scriptsize{}0} & {\scriptsize{}-1} & {\scriptsize{}0} & {\scriptsize{}+1} & {\scriptsize{}+1} & {\scriptsize{}0} & {\scriptsize{}-1}\tabularnewline
\hline 
{\scriptsize{}e} & {\scriptsize{}1} & {\scriptsize{}c} & {\scriptsize{}=} & {\scriptsize{}-1} & {\scriptsize{}+1} & {\scriptsize{}-1} & {\scriptsize{}-2} & {\scriptsize{}-1} & {\scriptsize{}0} & {\scriptsize{}+2} & {\scriptsize{}+2}\tabularnewline
\hline 
{\scriptsize{}e} & {\scriptsize{}2} & {\scriptsize{}f} & {\scriptsize{}=} & {\scriptsize{}0} & {\scriptsize{}0} & {\scriptsize{}-1} & {\scriptsize{}0} & {\scriptsize{}+1} & {\scriptsize{}+1} & {\scriptsize{}0} & {\scriptsize{}-1}\tabularnewline
\hline 
{\scriptsize{}e} & {\scriptsize{}2} & {\scriptsize{}c} & {\scriptsize{}=} & {\scriptsize{}-1} & {\scriptsize{}+1} & {\scriptsize{}0} & {\scriptsize{}-2} & {\scriptsize{}-2} & {\scriptsize{}+1} & {\scriptsize{}+2} & {\scriptsize{}+1}\tabularnewline
\hline 
{\scriptsize{}e} & {\scriptsize{}3} & {\scriptsize{}f} & {\scriptsize{}=} & {\scriptsize{}-1} & {\scriptsize{}+1} & {\scriptsize{}-2} & {\scriptsize{}-1} & {\scriptsize{}+2} & {\scriptsize{}+2} & {\scriptsize{}0} & {\scriptsize{}-1}\tabularnewline
\hline 
{\scriptsize{}e} & {\scriptsize{}3} & {\scriptsize{}c} & {\scriptsize{}=} & {\scriptsize{}0} & {\scriptsize{}0} & {\scriptsize{}0} & {\scriptsize{}-1} & {\scriptsize{}-1} & {\scriptsize{}0} & {\scriptsize{}+1} & {\scriptsize{}+1}\tabularnewline
\hline 
\end{tabular}%
\begin{tabular}{|c|c|c|c|c|c|c|c|c|c|c|c|}
\hline 
\multicolumn{12}{|c|}{\textbf{\scriptsize{}Enlaces}}\tabularnewline
\hline 
\multicolumn{4}{|c|}{\textbf{\scriptsize{}Vecinos}} & \multicolumn{2}{c|}{\textbf{\scriptsize{}1NN}} & \multicolumn{6}{c|}{\textbf{\scriptsize{}2NN}}\tabularnewline
\hline 
{\scriptsize{}e} & {\scriptsize{}4} & {\scriptsize{}f} & {\scriptsize{}=} & {\scriptsize{}+1} & {\scriptsize{}-1} & {\scriptsize{}-1} & {\scriptsize{}0} & {\scriptsize{}+2} & {\scriptsize{}+2} & {\scriptsize{}-1} & {\scriptsize{}-2}\tabularnewline
\hline 
{\scriptsize{}e} & {\scriptsize{}4} & {\scriptsize{}c} & {\scriptsize{}=} & {\scriptsize{}-1} & {\scriptsize{}+1} & {\scriptsize{}0} & {\scriptsize{}-1} & {\scriptsize{}-2} & {\scriptsize{}-1} & {\scriptsize{}+2} & {\scriptsize{}+2}\tabularnewline
\hline 
{\scriptsize{}e} & {\scriptsize{}5} & {\scriptsize{}f} & {\scriptsize{}=} & {\scriptsize{}-1} & {\scriptsize{}+1} & {\scriptsize{}-2} & {\scriptsize{}0} & {\scriptsize{}+1} & {\scriptsize{}+2} & {\scriptsize{}+1} & {\scriptsize{}-2}\tabularnewline
\hline 
{\scriptsize{}e} & {\scriptsize{}5} & {\scriptsize{}c} & {\scriptsize{}=} & {\scriptsize{}0} & {\scriptsize{}0} & {\scriptsize{}0} & {\scriptsize{}-1} & {\scriptsize{}-1} & {\scriptsize{}0} & {\scriptsize{}+1} & {\scriptsize{}+1}\tabularnewline
\hline 
{\scriptsize{}e} & {\scriptsize{}6} & {\scriptsize{}f} & {\scriptsize{}=} & {\scriptsize{}+1} & {\scriptsize{}-1} & {\scriptsize{}+1} & {\scriptsize{}+2} & {\scriptsize{}+1} & {\scriptsize{}0} & {\scriptsize{}-2} & {\scriptsize{}-2}\tabularnewline
\hline 
{\scriptsize{}e} & {\scriptsize{}6} & {\scriptsize{}c} & {\scriptsize{}=} & {\scriptsize{}-1} & {\scriptsize{}+1} & {\scriptsize{}-2} & {\scriptsize{}-2} & {\scriptsize{}0} & {\scriptsize{}+1} & {\scriptsize{}+2} & {\scriptsize{}+1}\tabularnewline
\hline 
\multicolumn{12}{|c|}{\textbf{\scriptsize{}Descripción de la CU Alternativa}}\tabularnewline
\hline 
\multicolumn{6}{|c|}{\textbf{\scriptsize{}NFCU}} & \multicolumn{6}{c|}{{\scriptsize{}4}}\tabularnewline
\hline 
\multicolumn{6}{|c|}{\textbf{\scriptsize{}NCCU}} & \multicolumn{6}{c|}{{\scriptsize{}4}}\tabularnewline
\hline 
\multicolumn{6}{|c|}{\textbf{\scriptsize{}NsCU}} & \multicolumn{6}{c|}{{\scriptsize{}3}}\tabularnewline
\hline 
\multicolumn{6}{|c|}{\textbf{\scriptsize{}NeCU}} & \multicolumn{6}{c|}{{\scriptsize{}6}}\tabularnewline
\hline 
\multicolumn{6}{|c|}{\textbf{\scriptsize{}ACU}} & \multicolumn{6}{c|}{{\scriptsize{}16}}\tabularnewline
\hline 
\multicolumn{6}{|c|}{\textbf{\scriptsize{}NhCU}} & \multicolumn{6}{c|}{{\scriptsize{}7}}\tabularnewline
\hline 
\multicolumn{6}{|c|}{\textbf{\scriptsize{}Z}} & \multicolumn{6}{c|}{{\scriptsize{}4}}\tabularnewline
\hline 
\end{tabular}\caption{\textbf{\footnotesize{}\label{fig:Hoja de datos (3.6.3.6)}}\textbf{Tabla
Superior. }Se muestra el tipo de vértice que da origen a la red, la
CU estándar, y la CU alternativa propuesta en este trabajo.\textbf{
Tablas inferiores.} Dado el i-ésimo sitio de la CU (si), los s i f
en el rango 1NN son las f-coordenadas de los enlaces primeros vecinos
de si, en el rango 2NN son las f-coordenadas de los sitios segundos
vecinos de si; análogamente, s i c en el rango 1NN son las c-coordenadas
de los enlaces primeros vecinos de si, en el rango 2NN son las c-coordenadas
de los sitios segundos vecinos de si; 'f' se refiere a filas, y 'c'
a columnas. \protect \\
Una explicación similar a la dada para el i-ésimo sitio, si, es válida
para el i-ésimo enlace, ei. los e i f en el rango 1NN son las f-coordenadas
de los sitios primeros vecinos de ei, en el rango 2NN son las f-coordenadas
de los enlaces segundos vecinos de ei; análogamente, los e i c en
el rango 1NN son las c-coordenadas de los sitios primeros vecinos
de ei, en el rango 2NN son las c-coordenadas de los enlaces segundos
vecinos de ei. \protect \\
Las coordenadas (f,c) de los primeros y segundos vecinos de un dado
sitio (enlace), se miden respecto el sitio (enlace) en cuestión. También
se muestran para la CU alternativa, el número de filas de la CU (NFCU),
el número de columnas de la CU (NCCU), el número de sitios de la CU
(NsCU), el número de enlaces de la CU (NeCU), área de la CU (ACU),
el número de huecos de la CU (NhCU), y el grado de cada vértice de
la red (Z).}
\end{figure}

Para completar la caracterización de los sitios y enlaces de la red
de kagome o (3.6.3.6), presentamos en la Figura \ref{fig:Hoja de datos (3.6.3.6)},
una hoja de datos con la siguiente información:
\begin{itemize}
\item \textbf{Tabla Superior. }Se muestra el tipo de vértice que da origen
a la red, la CU estándar, y la CU alternativa propuesta en este trabajo.\textbf{ }
\item \textbf{Tablas inferiores.} Dado el i-ésimo sitio de la CU (si), los
s i f en el rango 1NN son las f-coordenadas de los enlaces primeros
vecinos de si, en el rango 2NN son las f-coordenadas de los sitios
segundos vecinos de si; análogamente, s i c en el rango 1NN son las
c-coordenadas de los enlaces primeros vecinos de si, en el rango 2NN
son las c-coordenadas de los sitios segundos vecinos de si; 'f' se
refiere a filas, y 'c' a columnas. \\
Una explicación similar a la dada para el i-ésimo sitio, si, es válida
para el i-ésimo enlace, ei. los e i f en el rango 1NN son las f-coordenadas
de los sitios primeros vecinos de ei, en el rango 2NN son las f-coordenadas
de los enlaces segundos vecinos de ei; análogamente, los e i c en
el rango 1NN son las c-coordenadas de los sitios primeros vecinos
de ei, en el rango 2NN son las c-coordenadas de los enlaces segundos
vecinos de ei. \\
Las coordenadas (f,c) de los primeros y segundos vecinos de un dado
sitio (enlace), se miden respecto el sitio (enlace) en cuestión. También
se muestran para la CU alternativa, el número de filas de la CU (NFCU),
el número de columnas de la CU (NCCU), el número de sitios de la CU
(NsCU), el número de enlaces de la CU (NeCU), área de la CU (ACU),
el número de huecos de la CU (NhCU), y el grado de cada vértice de
la red (Z).
\end{itemize}
Una hoja de datos similar se puede implementar para cada una de las
redes Arquimedianas. Los vectores desplazamientos hacia los primeros
y segundos vecinos de un dado sitio o enlace (f,c), son fundamentales
para el recorrido de un cluster de sitios conectados, y la posterior
determinación de la condición de percolación de la red.

\section{Curvas de fase de las redes Arquimedianas por deposición de monómeros}

\subsection{Preliminares}

En esta sección mostramos en detalle una manera de obtener las curvas
de fase de cada una de las redes Arquimedianas a partir de la representación
alternativa explicada en la sección anterior; lo haremos tomando como
ejemplo la red de kagome o (3.6.3.6). Dado que la CU asociada a cada
red Arquimediana es cuadrada o rectangular, podemos trasladar esta
CU horizontalmente o verticalmente las veces que sea necesario, hasta
lograr que las aristas de la red sean del tamaño deseado, logrando
a su vez que la misma también sea cuadrangular o rectangular. En nuestro
caso consideramos redes cuadrangulares. Para cualquiera de las redes
Arquimedianas, trabajamos con aristas de tamaño 96, 192, 364, y 768
(todas múltiplos de ¡48!). Como vimos en la Sección \ref{subsec:Etiquetado-y-caracterizaci=0000F3n},
la cuadratura de cada una de las redes Arquimedianas, nos permite
identificar o etiquetar a cada celda cuadrada que compone una red
dada, con el par ordenado de números enteros $(f_{i},c_{i})$, donde
$f_{i}$ es la i-ésima fila de la red contada de arriba hacia abajo,
y $c_{i}$ es la i-ésima columna de red contada de izquierda a derecha,
como sucede cuando identificamos los elementos de una matriz de números
en el sentido matemático (ver Tabla \ref{tab:Caracterizaci=0000F3n de los sitios y enlaces de la CU de la red (3.6.3.6)}).
También, la caracterización de cada sitio y enlace de la misma lo
hicimos tal como se describe en dicha sección.

Utilizamos el lenguaje de programación C++ \citep{sedgewick1995algoritmos}
en las simulaciones computacionales para la obtención de las curvas
de fase.

\subsection{Proceso de ocupación aleatoria de sitios y enlaces de las redes Arquimedianas}

El llenado aleatorio de cada red hasta las concentraciones deseadas
de sitios y enlaces, lo hicimos siguiendo los lineamientos descriptos
por Efros et al \citep{efros1982physics}.

Denotemos con $G$ una grafo o red cualquiera, con $S(G)$ su conjunto
de sitios, y con $B(G)$ su conjunto de enlaces. Asignamos aleatoriamente
a cada sitio de $S$ y a cada enlace $B$, un número entre cero (0)
y uno (1) . Llamamos $P=\left\{ p_{1},p_{2},...\right\} $ al conjunto
de números asignados a $S$, y $W=\left\{ w_{1},w_{2},...\right\} $
al conjunto de números asignados a $B$. Si $p$ es la probabilidad
con la que queremos ocupar cada sitio de $S$, entonces todo i-ésimo
sitio $s_{i}$ de este, cuyo $p_{i}$ asignado en $P$ verifique la
desigualdad $p_{i}<=p$ , será ocupado con probabilidad constante
$p$. De la misma manera, si $w$ es la probabilidad con la que queremos
ocupar cada enlace de $B$, todo j-ésimo enlace $b_{j}$ de este,
cuyo $w_{j}$ asignado en $W$ verifique la desigualdad $w_{j}<=w$,
será ocupado con probabilidad constante $w$. 

\subsection{Proceso de obtención de la curva de fase de la red (3.6.3.6)}

\subsubsection{Encapsulamiento de la hoja de datos de la red (3.6.3.6)}

El núcleo del algoritmo utilizado para la obtención de la curva de
fase lo constituye la clase \_3C6C3C6. Esta consta esencialmente de
dos partes. En su primera parte se almacenan los datos de la red obtenidos
por visualización directa de la misma y de su CU, y además se obtienen
otros datos auxiliares a través de su constructor \_3C6C3C6(). La
segunda parte esta constituida por los métodos o funciones, que manipulan
los datos de la primera parte, para obtener la curva de fase. 

En el Algoritmo \ref{alg:primera parte de la clase _(3.6.3.6)}, se
muestra la primera parte de la clase \_3C6C3C6 en donde se encapsulan
gran parte de los datos contenidos en la hoja de datos mostrada en
la Figura \ref{fig:Hoja de datos (3.6.3.6)}. Los arreglos NFR{[}{]}
y NCR{[}{]} contienen el número de filas, y el número de columnas
de cada una de las redes consideradas, respectivamente. NFCU y NCCU,
representan el número de filas, y el número de columnas de la CU,
respectivamente. Mientras que NsCU y NeCU, representan el número de
sitios y el número de enlaces contenidos en la CU, respectivamente.
La variable 'enlaces\_iterados', representa el número de probabilidades
$P_{b}$ con que se depositaron los enlaces en cada red considerada.
NCUV{[}{]}, y NCUH{[}{]}, representan el número de CU que caben en
una columna, y fila de la red, respectivamente, para cada tamaño de
red considerado. NUM\_CELDAS{[}{]} almacena el número total de CU
contenido en la red, para cada tamaño de red considerado. El arreglo
paso\_ITER\_ENLACES{[}{]}, almacena el paso con que irá cambiando
la concentración de enlaces o la probabilidad de deposición de los
enlaces, para cada tamaño de red considerado. NUM\_ENLACES{[}{]} y
NUM\_SITIOS{[}{]}, almacena el número total de enlaces, y de sitios
de la red, para cada tamaño de red considerado.

\subsubsection{Funciones o métodos que componen la clase \_3C6C3C6}

En el Algoritmo \ref{alg:segunda parte de _3C6C3C6}, se muestra la
segunda parte de la cual está compuesta la clase \_3C6C3C6. Todos
los métodos o funciones que la componen están realizadas de tal manera
que, el conjunto de parámetros introducido a través de sus argumentos,
es lo único que necesita para su funcionamiento. Lo que hace cada
función está descrito en el algoritmo con texto de color magenta.
En las subsecciones siguientes describiremos cómo hacen su trabajo
las funciones principales del algoritmo.

\subsubsection{Funciones \textcolor{blue}{llenoSitiosRed(...), llenoEnlacesRed(...),
}\textcolor{black}{y}\textcolor{blue}{{} llenoRed(...)}}

\paragraph{llenoSitiosRed():}

Como su propio nombre lo sugiere, esta función ocupa con monómeros
los sitios de la $Red$ con probabilidad \textcolor{blue}{$cubriS$.}
Identificado un sitio $(f,c)$ de la red, este será ocupado, $Red(f,c)=1$,
si se verifica la desigualdad, $V(f,c)<=cubriS$, sino se mantiene
desocupado, $Red(f,c)=0$. Los tres tipo de sitios que componen la
red, $s_{i}$, se identifican por las condiciones que deben cumplir
sus coordenadas $(f,c)$ respecto de la función módulo, tal como se
muestra en la Tabla \ref{tab:Caracterizaci=0000F3n de los sitios y enlaces de la CU de la red (3.6.3.6)}.

\paragraph{llenoEnlacesRed():}

Como su propio nombre lo sugiere, esta función ocupa con monómeros
los enlaces de la $Red$ con probabilidad \textcolor{blue}{$cubriE$.}
Identificado un enlace $(f,c)$ de la red, este será ocupado, $Red(f,c)=1$,
si se verifica la desigualdad, $V(f,c)<=cubriE$, sino se mantiene
desocupado, $Red(f,c)=0$. Los seis tipo de enlaces que componen la
red, $b_{i}$, se identifican por las condiciones que deben cumplir
sus coordenadas $(f,c)$ respecto de la función módulo, tal como se
muestra en la Tabla \ref{tab:Caracterizaci=0000F3n de los sitios y enlaces de la CU de la red (3.6.3.6)}.

\paragraph{llenoRed():}

Esta función ocupa los sitios de la red con probabilidad cubriS, y
los enlaces de la red con probabilidad cubriE. Hace uso directo de
las funciones \textit{llenoSitiosRed()}, y \textit{llenoEnlacesRed()}.

\subsubsection{\textcolor{black}{Función }\textcolor{blue}{Ps\_IM(...)}}

El Algoritmo \ref{alg:Algoritmo de bisecci=0000F3n}, muestra el algoritmo
de bisección seguido por la función Ps\_IM() para determinar la probabilidad
de Hammersley para una dada concentración de enlaces $P_{b}$, dentro
de una precisión dada por la variable $PRECISION$. Este algoritmo
es el mismo que se utiliza para encontrar las raíces de una función
$f(x)$ dentro del intervalo $[a.b]$.

\subsubsection{Función \textcolor{blue}{Psi(...)}}

El Algoritmo \ref{alg:Algoritmo que usa <Psi>} muestra como se implementa
la función Psi() para encontrar un promedio de la probabilidad de
Hammersley para una dada concentración de enlaces Pb. NRP es el número
de redes promediadas para obtener tal promedio, el cual es también
el número de veces que se actualiza la matriz, V, con números pertenecientes
al intervalo $[0,1]$. Es decir, el promedio de la probabilidad de
Hammersley, surge de considerar NRP configuraciones de la concentración
Pb de enlaces depositados. Al actualizar aleatoriamente los valores
de V, también las configuraciones de sitios ocupados se actualizan.

\subsubsection{Función \textcolor{blue}{PsPromYESTD(...)}}

El Algoritmo \ref{alg:Algoritmo PsPromYESTD} muestra cómo se implementa
la función \textcolor{black}{PsPromYESTD()} para encontrar el valor
final de la probabilidad de Hammersley con su error estándar ESTD,
para una dada concentración de enlaces Pb. NPsi es el número de Psi()
almacenados en el arreglo psi{[}{]}, que luego será usado para encontrar
el valor final de la probabilidad de Hammersley, Ps\_prom, con su
respectivo error estándar ESTD.

\subsubsection{Función\textcolor{blue}{{} main()}}

El Algoritmo \ref{alg:AlgoritmoMain} muestra cómo se implementó la
función \textcolor{black}{main()} para encontrar el valor final de
la probabilidad de Hammersley PsProm con su error estándar ESTD, para
cada una de las concentraciones de enlaces $P_{b}$ consideradas,
y para cada uno de los tamaños de red considerados. El proceso es
como sigue:
\begin{enumerate}
\item \textbf{MiGenRan gen}; La variable \textbf{gen} es una instancia de
la clase \textbf{MiGenRan} la cual define un generador random C++.
\textbf{gen} acepta en su argumento una semilla la cual en este trabajo
tomamos como la hora en segundos, contados a partir de 1970. La semilla
se mantiene constante durante toda la ejecución de este algoritmo.
\item \textbf{\_3C63C6 R}; Define la instancia \textbf{R} de la clase \textbf{\_3C63C6. }
\item \textbf{MODELO} es la variable mediante la cual se especifica el modelo
a simular. Especificamos el modelo $S\cup B$ mediante la asignación
MODELO = 'OR'; y el modelo $S\cap B$ mediante la asignación MODELO=
'AND'; Si MODELO='OR', entonces ks\_max = R.n12s y ke\_max = R.n12e;
sino si MODELO = 'AND', entonces ks\_max = R.n1s y ke\_max = R.n1e.
Recordemos que las variables n12s, n12e, n1s, y n1e, se definen en
la clase \textbf{\_3C63C6}. n12s es el número de primeros vecinos
más el número de segundos vecinos de un dado sitio; n12e es el número
de primeros vecinos más el número de segundos vecinos de un dado enlace;
n1s es el número de primeros vecinos de un dado sitio y, n1e es el
número de primeros vecinos de un dado enlace. Por lo tanto, consideramos
al modelo $S\cup B$ como una interacción hasta segundos vecinos de
sitios y enlaces; mientras que al modelo $S\cap B$ lo consideramos
como una interacción entre primeros vecinos de sitios y enlaces.
\item \textbf{AbreArchivo()}; Esta función abre un archivo de datos cuyo
nombre se define como una variable global del programa. El archivo
almacenará los tamaños de red considerados, y los umbrales críticos
de sitios para cada concentración de enlaces considerada. En la siguiente
sección mostramos su estructura.
\item \textbf{PsProm} y \textbf{ESTD}. PsProm es un doble promedio de los
umbrales críticos de sitios, para un dado tamaño de red, y para una
concentración de enlaces dada. ESTD es el error estándar con que se
mide PsProm.
\item \textbf{iter\_red} itera sobre el número de tamaños de aristas de
red considerados. En este trabajo iteramos sobre cuatro tamaños de
aristas, que son: 96, 192, 384 y 768.
\item \textbf{nfr} y \textbf{ncr} representan el número de filas y columnas
de la red considerada, respectivamente; nS y nE representan el número
de sitios y enlaces de la red considerada, respectivamente; pasoE
es el paso con que se itera la concentración de enlaces $P_{b}$,
hasta cubrir el número de enlaces de la red, nE.
\item Según la red Arquimediana considerada, y el modelo de interacción
considerado ($S\cup B$ o $S\cap B$), puede ser necesario redefinir
los bordes de la red. Se necesitan cuatro variables para definir los
bordes de una red, dos para los bordes horizontales, y dos para los
bordes verticales.
\item \textbf{iter\_pb} itera sobre todas las concentraciones de enlaces
$P_{b}$ consideradas, las cuales están definidas por la variable
\textbf{'enlaces\_iterados'} en la clase \_3C6C3C6.
\item \textbf{R.PsPromYESTD()} devuelve el umbral crítico de sitios \textbf{PsProm}
y su error estándar \textbf{ESTD}, para cada tamaño de red y concentración
de enlaces $P_{b}$ considerados, los cuales se irán almacenando en
el archivo de datos abierto en el ítem 4.
\end{enumerate}
\begin{algorithm}
\textcolor{blue}{class} \textbf{\_3C6C3C6} \textcolor{blue}{\{}

\textcolor{blue}{public:}

\textcolor{magenta}{// Primeros y segundos vecinos de los sitios que
componen la celda unidad.}

\textcolor{blue}{static const int n1s=4, n2s=4, n12s=n1s+n2s;}

\textcolor{blue}{int s1\_f{[}n12s{]}= \{-1, \textvisiblespace{}0,+1,
0,\hspace{5mm}-2, 0, +2, 0 \};}

\textcolor{blue}{int s1\_c{[}n12s{]}=\{ 0, -1, 0,+1,\hspace{5mm}
0,-2, 0, +2 \};}

\textcolor{blue}{int s2\_f{[}n12s{]}= \{ 0,+1, 0, -1,\hspace{5mm}
0,+2, 0, -2 \};}

\textcolor{blue}{int s2\_c{[}n12s{]}=\{-1,-1,+1,+1,\hspace{5mm}-2,-2,+2,+2\};}

\textcolor{blue}{int s3\_f{[}n12s{]}= \{-1,+1,+1,-1,\hspace{5mm}-2,+2,+2,-2\};}

\textcolor{blue}{int s3\_c{[}n12s{]}=\{ 0, -1, 0,+1,\hspace{5mm}
0, -2, 0, +2 \};}

\textcolor{magenta}{// Primeros y segundos vecinos de los enlaces}
\textcolor{magenta}{que componen la celda unidad.}

\textcolor{blue}{static const int n1e=2, n2e=6, n12e=n1e+n2e;}

\textcolor{blue}{int e1\_f{[}n12e{]}=\{ 0, 0, \hspace{5mm}-1, 0,
+1, +1, 0, -1 \};}

\textcolor{blue}{int e1\_c{[}n12e{]}=\{-1,+1, \hspace{5mm}-1,-2,-1,
0,+2,+2 \};}

\textcolor{blue}{int e2\_f{[}n12e{]}=\{ 0, 0,\hspace{5mm} -1, 0,
+1, +1, 0, -1 \};}

\textcolor{blue}{int e2\_c{[}n12e{]}=\{-1,+1,\hspace{5mm} 0,-2,-2,+1,+2,
+1 \};}

\textcolor{blue}{int e3\_f{[}n12e{]}=\{-1,+1, \hspace{5mm}-2,-1,+2,
+2, 0, -1 \};}

\textcolor{blue}{int e3\_c{[}n12e{]}=\{ 0, 0, \hspace{5mm}0, -1,
-1, 0, +1, +1 \};}

\textcolor{blue}{int e4\_f{[}n12e{]}=\{+1,-1, \hspace{5mm}-1, 0,+2,+2,-1,
-2 \};}

\textcolor{blue}{int e4\_c{[}n12e{]}=\{-1,+1, \hspace{5mm}0,-1,-2,
-1,+2,+2 \};}

\textcolor{blue}{int e5\_f{[}n12e{]}=\{-1,+1, \hspace{5mm}-2, 0,+1,+2,+1,-2
\};}

\textcolor{blue}{int e5\_c{[}n12e{]}=\{ 0, 0,\hspace{5mm} 0,-1, -1,
0, +1, +1 \};}

\textcolor{blue}{int e6\_f{[}n12e{]}=\{+1,-1, \hspace{5mm}+1,+2,+1,
0,-2,-2 \};}

\textcolor{blue}{int e6\_c{[}n12e{]}=\{-1,+1,\hspace{5mm} -2,-2,
0,+1,+2,+1\};}

\textcolor{blue}{public: static const int NUMREDES = 4;}

\textcolor{blue}{int NFR{[}NUMREDES{]} = \{96, 192, 384, 768\}, NCR{[}NUMREDES{]}
= \{96, 192, 384, 768\};}

\textcolor{blue}{int NFCU=4, NCCU=4,NsCU=3,NeCU=6, enlaces\_iterados
= 48;}

\textcolor{magenta}{// Otras variables}

\textcolor{blue}{int NCUV{[}NUMREDES{]}, NCUH{[}NUMREDES{]};}

\textcolor{blue}{int NUM\_CELDAS{[}NUMREDES{]}, pasoITER\_ENLACES{[}NUMREDES{]};}

\textcolor{blue}{int NUM\_ENLACES{[}NUMREDES{]}, NUM\_SITIOS{[}NUMREDES{]};}

\textcolor{magenta}{// Constructor}

\textcolor{blue}{\_3C6C3C6() \{ InicializoArreglosDatosRedes(); \}}

\textcolor{magenta}{// Segunda parte no mostrada.}

.

.

.

\textcolor{blue}{\}}

\caption{\label{alg:primera parte de la clase _(3.6.3.6)}Primera parte de
la clase \textbf{\_3C6C3C6 }que muestra el encapsulamiento de los
datos contenidos en la hoja de datos de la red. Donde: n1s=número
de primeros vecinos de un sitio; n2s=número de segundos vecinos de
un sitio; n12s=n1s+n2s; n1e=número de primeros vecinos de un enlace;
n2e=número de segundos vecinos de un enlace; n12e=n1e+n2e. En un arreglo
cualquiera s1f{[}{]} o s1c{[}{]}, los primeros n1s elementos son primeros
vecinos, y los n2s restantes son segundos vecinos. Similarmente, en
un arreglo cualquiera e1f{[}{]} o e1c{[}{]}, los primeros n1e elementos
son primeros vecinos, y los n2e restantes son segundos vecinos.}
\end{algorithm}

\pagebreak{}

\begin{algorithm}
\textcolor{blue}{class} \textbf{\_3C6C3C6} \textcolor{blue}{\{}

\textcolor{blue}{public:}

\textcolor{magenta}{// Primera parte no mostrada}

.

.

.

\textcolor{magenta}{// Segunda parte}

\textcolor{magenta}{// Obtiene datos auxiliares de la primera parte:}

\textcolor{blue}{void InicializoArreglosDatosRedes();}

\textcolor{magenta}{// Ocupa los sitios de la Red con probabilidad
cubriS:}

\textcolor{blue}{void llenoSitiosRed (double{*}{*} V, bool{*}{*} Red,
double cubriS, int NFR, int NCR, int NFCU, int NCCU);}

\textcolor{magenta}{// Ocupa los enlaces de la Red con probabilidad
cubriE:}

\textcolor{blue}{void llenoEnlacesRed(double{*}{*} V, bool{*}{*} Red,
double cubriE, int NFR, int NCR, int NFCU, int NCCU);}

\textcolor{magenta}{// Ocupa los sitios y enlaces de la Red con probabilidad
cubriE y cubriS, respectivamente:}

\textcolor{blue}{void llenoRed (double{*}{*} V, bool{*}{*} Red, double
cubriE, double cubriS, int NFR, int NCR, int NFCU, int NCCU);}

\textcolor{magenta}{// Verifica que el sitio o enlace (fil,col) de
la red esté ocupado, que no haya sido visitado,}

\textcolor{magenta}{// y que sus coordenadas no desborden la red.
Devuelve 1 si todas estas condiciones se verifican :}

\textcolor{blue}{bool DesbordeOcupadoVisitado (bool{*}{*} M,int fil,int
col, bool{*}{*} visited,int NFR,int NCR);}

\textcolor{magenta}{// Recorre todos los sitios y enlaces ocupados
de un dado cluster, a partir un nodo raíz dado por}

\textcolor{magenta}{// la función percoRightDown\_CCA (). Si el sitio
o enlace (fil,col) pertenece al borde superior}

\textcolor{magenta}{// hará flagfil1=true, y si pertenece al borde
inferior hará flagfil2=true. Similarmente,}

\textcolor{magenta}{// si el sitio o enlace (fil,col) pertenece al
borde izquierdo hará flagcol1=true,}

\textcolor{magenta}{// y si pertenece al borde derecho hará flagcol2=true:}

\textcolor{blue}{void DFS\_CCA\_conflag (bool{*}{*} M, int row, int
col, bool{*}{*} visited, int NFR, int NCR, bool \&flagfil1, bool \&flagfil2,
bool \&flagcol1, bool \&flagcol2, int n12s, int n12e);}

\textcolor{magenta}{// Auxilia a DFS\_CCA\_conflag() en la compresión
del código:}

\textcolor{blue}{void auxiliar (bool{*}{*} M,bool{*}{*} visited,int
row,int col, int NFR,int NCR,int ks\_max, int ke\_max, bool \&flagfil1,bool
\&flagfil2, bool \&flagcol1,}

\textcolor{blue}{bool \&flagcol2, int bordefil1,int bordefil2, int
bordecol1,int bordecol2, int{*} soe\_f, int{*} soe\_c,int ksoe\_max);}

\textcolor{magenta}{// Detecta en la Red un primer sitio o enlace
ocupado (nodo raíz), y se lo pasa a} \textcolor{magenta}{DFS\_CCA\_conflag().}

\textcolor{magenta}{// Luego cuando esta última haya recorrido todo
el cluster, chequea si la red ha percolado revisando los valores}

\textcolor{magenta}{// de flagfil1, flagfil2, flagcol1, flagcol2.
Si la Red percola hace pr=1 o pd=1, entregando luego el control al
resto del programa,}

\textcolor{magenta}{// sino busca otro nodo raíz:}

\textcolor{blue}{void percoRightDown\_CCA (bool{*}{*} M,int bordefil1,int
bordefil2, int bordecol1,int bordecol2, int NFR,int NCR,int \&pr,int
\&pd, int ks\_max, int ke\_max);}

\textcolor{magenta}{// Para un dado Pb, devuelve la probabilidad de
Hammersley Ps por el método del semi intervalo:}

\textcolor{blue}{double Ps\_IM (double{*}{*} V,bool{*}{*} Red,double
Pb, double cubriSi, double cubriSf, double PRECISION, int nfr, int
ncr, int NFCU, int NCCU,}

\textcolor{blue}{int bordefil1,int bordefil2, int bordecol1,int bordecol2,
int ks\_max, int ke\_max);}

\textcolor{magenta}{// Para una dada probabilidad Pb, encuentra un
valor promedio de la probabilidad de Hammersley, Ps}

\textcolor{magenta}{// para la cual la red percola en alguna dirección:}

\textcolor{blue}{double Psi (MiGenRan\& gen, double Pb, double cubriSi,
double cubriSf, double PRECISION, int nfr, int ncr, int NFCU, int
NCCU, int bordefil1,}

\textcolor{blue}{int bordefil2, int bordecol1,int bordecol2, int ks\_max,
int ke\_max);}

\textcolor{magenta}{// Para una dada probabilidad Pb, encuentra el
valor promedio PsProm, de los ya promediados valores devueltos}

\textcolor{magenta}{// por Psi(), con su error estándar ESTD:}

\textcolor{blue}{void PsPromYESTD (MiGenRan\& gen,double Pb,double
cubriSi, double cubriSf,double PRECISION,int nfr, int ncr,int NFCU,int
NCCU, int bordefil1,}

\textcolor{blue}{int bordefil2, int bordecol1,int bordecol2, int ks\_max,
int ke\_max, int NumPsi,double\& PsProm, double\& ESTD);}

\textcolor{blue}{\};}

\caption{\label{alg:segunda parte de _3C6C3C6}Funciones que componen la segunda
parte de la clase \_3C6C3C6.}
\end{algorithm}

\pagebreak{}

\begin{algorithm}
\begin{centering}
\includegraphics[height=17cm]{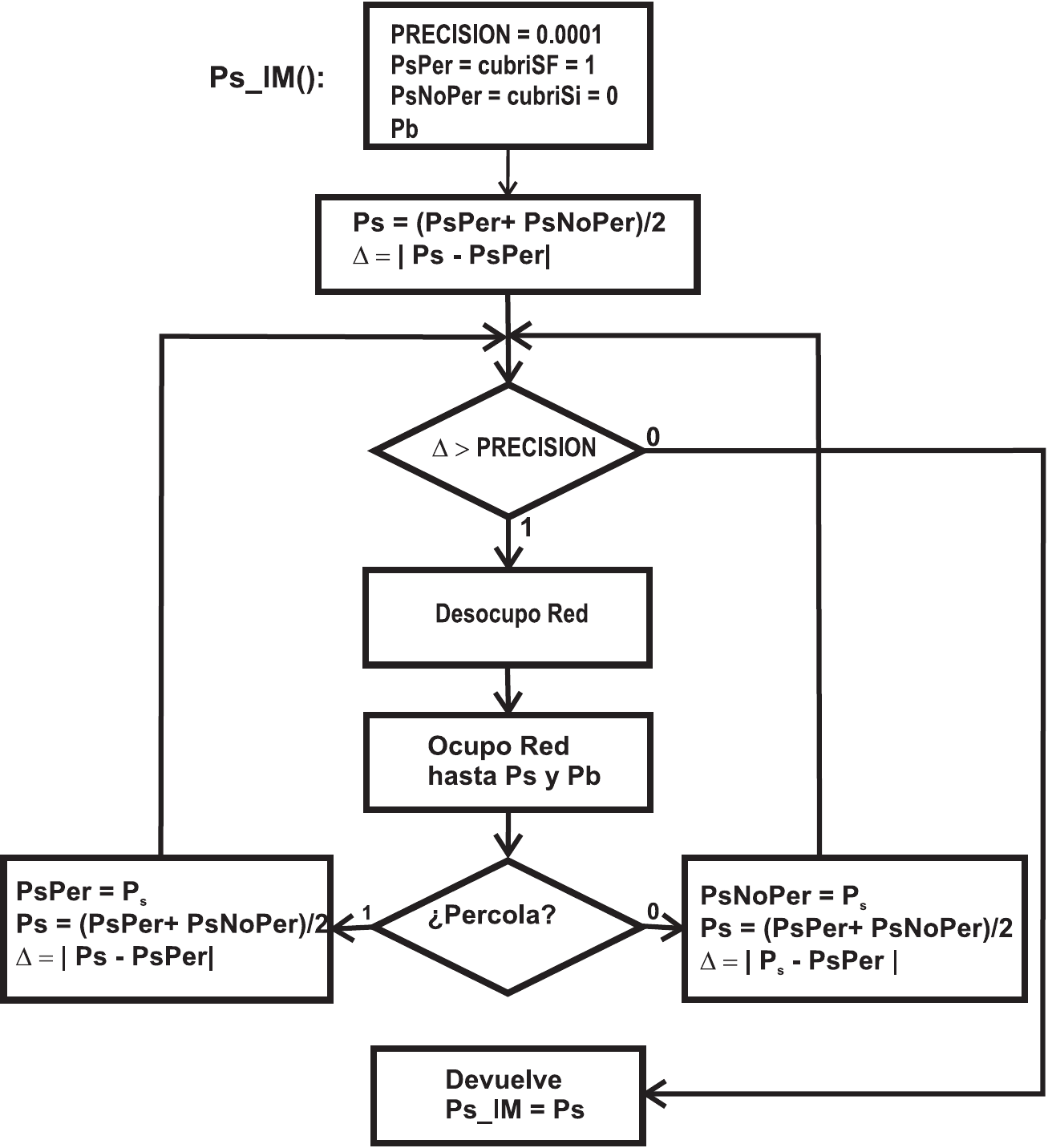}
\par\end{centering}
\caption{\label{alg:Algoritmo de bisecci=0000F3n}Algoritmo de bisección seguido
por la función Ps\_IM() para determinar la probabilidad de Hammersley
para una dada concentración de enlaces Pb, dentro de una precisión
dada por la variable PRECISION.}
\end{algorithm}

\pagebreak{}

\begin{algorithm}
\begin{centering}
\includegraphics[height=17cm]{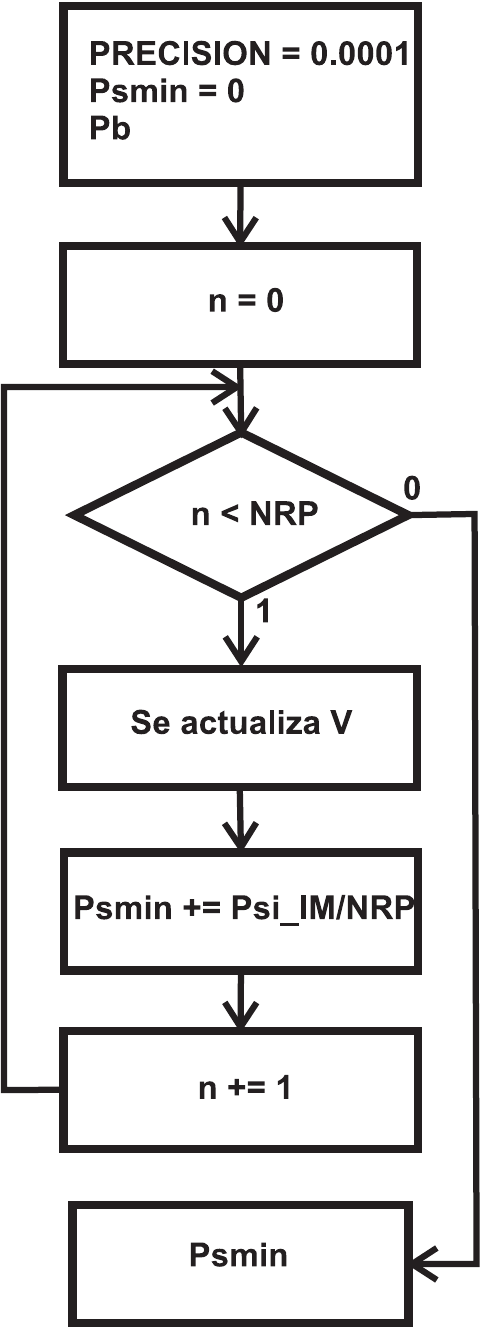}
\par\end{centering}
\caption{\label{alg:Algoritmo que usa <Psi>}Algoritmo que usa la función Psi()
para encontrar un promedio de la probabilidad de Hammersley. NRP es
el número de redes promediadas, el cual es también el número de veces
que se actualiza la matriz, V, con números pertenecientes al intervalo
{[}0,1{]}.}
\end{algorithm}

\pagebreak{}

\begin{algorithm}
\begin{centering}
\includegraphics[height=17cm]{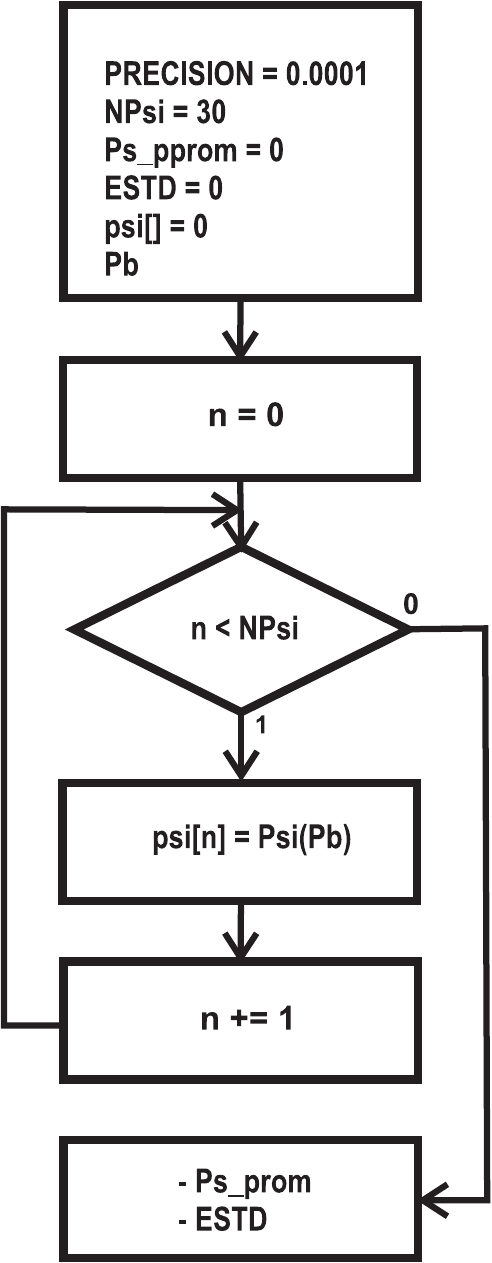}
\par\end{centering}
\caption{\label{alg:Algoritmo PsPromYESTD}Algoritmo que usa la función \textcolor{black}{PsPromYESTD()}
para encontrar el valor final de la probabilidad de Hammersley con
su error estándar ESTD, para una dada concentración de enlaces Pb.
NPsi es el número de Psi() almacenados en el arreglo psi{[}{]}, que
luego será usado para encontrar el valor final de la probabilidad
de Hammersley, Ps\_prom, con su respectivo error estándar ESTD.}
\end{algorithm}

\pagebreak{}

.
\begin{algorithm}
\begin{centering}
\includegraphics[height=16cm]{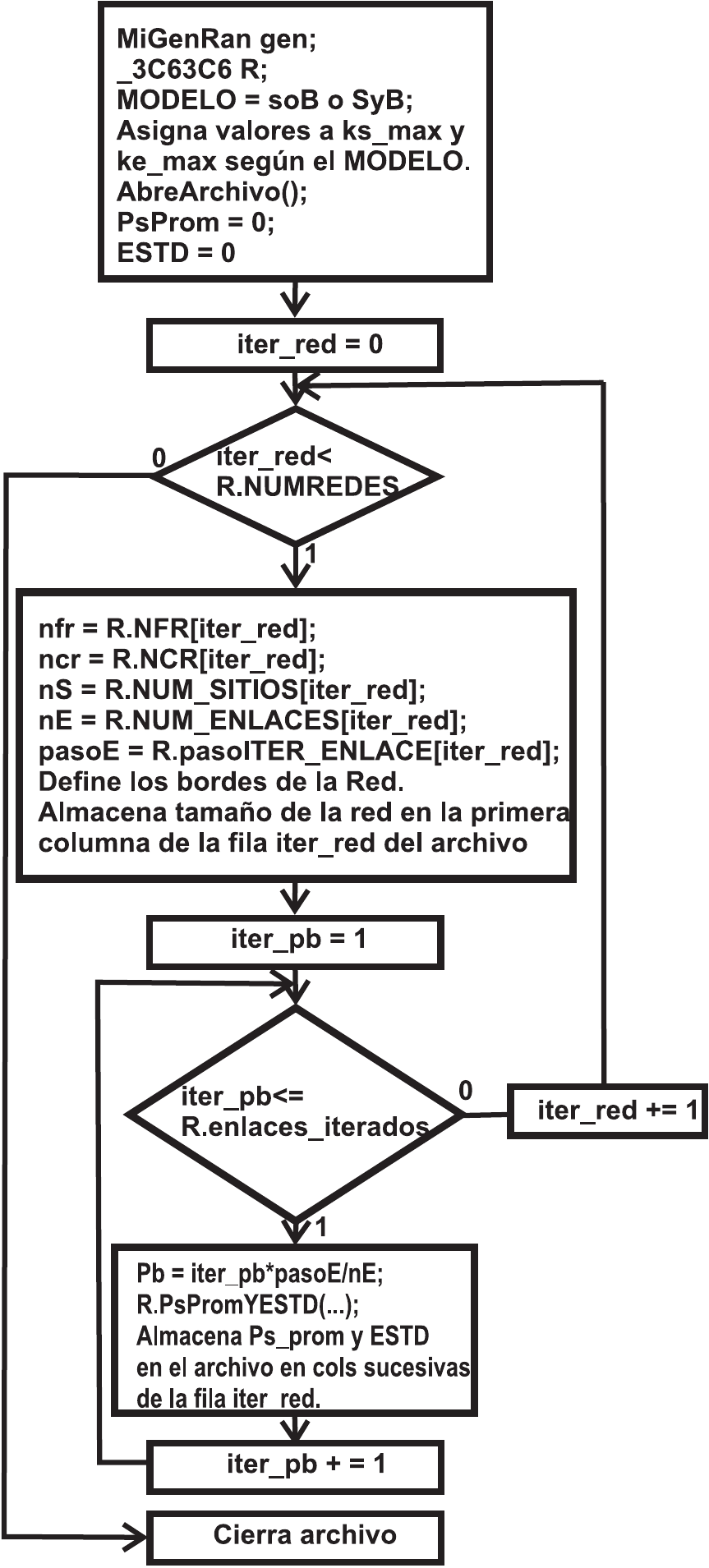}
\par\end{centering}
\caption{\label{alg:AlgoritmoMain}Algoritmo implementado en la función \textcolor{black}{main()}
para encontrar el valor final de la probabilidad de Hammersley con
su error estándar ESTD, para cada una de las concentraciones de enlaces
Pb y para cada uno de los tamaños de red considerados.}
\end{algorithm}

\subsection{Estructura de los datos obtenidos por main() para la red (3.6.6.3)
en el modelo S$\cup$\textmd{B}}

En la Tabla \ref{tab:extractodatoskagome} se muestra un extracto
del archivo de datos obtenido por la función \textbf{main()} a partir
del algoritmo \ref{alg:AlgoritmoMain} utilizado para obtener $P_{s}(L,P_{b})$
y $ESTD$, para cada tamaño de arista $L$, y para cada $P_{b}$ considerado
respecto de la red $(3.6.3.6)$. La primera y última fila se han agregado
aquí para clarificar la estructura del archivo. Como se ve desde la
última fila de esta tabla solo se muestran los $P_{s,c}(L,P_{b})$
y $ESTD$, correspondientes a los primeros seis valores de $P_{b}$,
de un total de 49 valores considerados. Como se observa en cada columna
$P_{s,c}(L,P_{b})$, sus valores dependen del tamaño de la red. 

\begin{table}
\begin{centering}
\begin{tabular}{|c|c|c|c|c|c|c|c|c|c|c|c|c|}
\hline 
\textbf{\footnotesize{}L} & \textbf{\footnotesize{}Ps,c} & \textbf{\footnotesize{}ESTD} & \textbf{\footnotesize{}Ps,c} & \textbf{\footnotesize{}ESTD} & \textbf{\footnotesize{}Ps,c} & \textbf{\footnotesize{}ESTD} & \textbf{\footnotesize{}Ps,c} & \textbf{\footnotesize{}ESTD} & \textbf{\footnotesize{}Ps,c} & \textbf{\footnotesize{}ESTD} & \textbf{\footnotesize{}Ps,c} & \textbf{\footnotesize{}ESTD}\tabularnewline
\hline 
\hline 
\textbf{\footnotesize{}96} & {\footnotesize{}0.6382} & {\footnotesize{}1.4E-4} & {\footnotesize{}0.6384} & {\footnotesize{}2.9E-4} & {\footnotesize{}0.6373} & {\footnotesize{}3.8E-4} & {\footnotesize{}0.6358} & {\footnotesize{}4.4E-4} & {\footnotesize{}0.6353} & {\footnotesize{}4.8E-4} & {\footnotesize{}0.6289} & {\footnotesize{}5.0E-4}\tabularnewline
\hline 
\textbf{\footnotesize{}192} & {\footnotesize{}0.644} & {\footnotesize{}4.6E-4} & {\footnotesize{}0.6438} & {\footnotesize{}4.6E-4} & {\footnotesize{}0.6426} & {\footnotesize{}4.6E-4} & {\footnotesize{}0.6420} & {\footnotesize{}4.7E-4} & {\footnotesize{}0.6388} & {\footnotesize{}4.7E-4} & {\footnotesize{}0.6352} & {\footnotesize{}4.6E-4}\tabularnewline
\hline 
\textbf{\footnotesize{}384} & {\footnotesize{}0.6479} & {\footnotesize{}4.5E-4} & {\footnotesize{}0.6473} & {\footnotesize{}4.5E-4} & {\footnotesize{}0.6463} & {\footnotesize{}4.5E-4} & {\footnotesize{}0.6448} & {\footnotesize{}4.6E-4} & {\footnotesize{}0.6422} & {\footnotesize{}4.5E-4} & {\footnotesize{}0.6386} & {\footnotesize{}4.5E-4}\tabularnewline
\hline 
\textbf{\footnotesize{}768} & {\footnotesize{}0.6500} & {\footnotesize{}4.5E-4} & {\footnotesize{}0.6496} & {\footnotesize{}4.5E-4} & {\footnotesize{}0.6482} & {\footnotesize{}4.5E-4} & {\footnotesize{}0.6468} & {\footnotesize{}4.5E-4} & {\footnotesize{}0.6443} & {\footnotesize{}4.5E-4} & {\footnotesize{}0.6405} & {\footnotesize{}4.5E-4}\tabularnewline
\hline 
\textbf{\footnotesize{}Pb} & \multicolumn{2}{c|}{{\footnotesize{}0/48}} & \multicolumn{2}{c|}{{\footnotesize{}1/48}} & \multicolumn{2}{c|}{{\footnotesize{}2/48}} & \multicolumn{2}{c|}{{\footnotesize{}3/48}} & \multicolumn{2}{c|}{{\footnotesize{}4/48}} & \multicolumn{2}{c|}{{\footnotesize{}5/48}}\tabularnewline
\hline 
\end{tabular}
\par\end{centering}
\caption{\label{tab:extractodatoskagome}Las filas etiquetadas con los tamaños
de red 96, 192, 384, y 768, constituyen un extracto del archivo de
datos obtenido por la función main() a partir del algoritmo \ref{alg:AlgoritmoMain}
utilizado para obtener Ps y ESTD, para cada tamaño de red y para cada
Pb considerado. La primera y última fila se han agregado aquí para
clarificar la estructura del archivo. Como se ve en la última fila,
en esta tabla solo se muestran los Ps,c y ESTD, correspondientes a
los primeros seis valores de Pb, de un total de 49 valores considerados. }
\end{table}

\subsection{Escaleo de los $P_{s,c}$ hacia $L\rightarrow\infty$: Curvas de
fase}

Una vez obtenidos los $P_{s,c}(L,P_{b})$, como los mostrados en la
Tabla \ref{tab:extractodatoskagome} para la red $(3.6.3.6)$, lo
que resta es extrapolar los resultados en cada columna $P_{s,c}(L,P_{b})$
hacia el límite termodinámico $L\rightarrow\infty$, lo cual se hace
con la relación de escaleo\citep{Stauffer1994}\\

\begin{equation}
P_{s,c}(L,P_{b})=P_{s,c}^{\infty}(P_{b})+AL^{-1/\nu}\label{eq:ecuacion de escaleo}
\end{equation}
\\

donde $P_{s,c}^{\infty}(P_{b})=lim_{L\rightarrow\infty}P_{s,c}(L,P_{b})$,
$A$ es una constante no universal, y $\nu=4/3$ es un exponente crítico
exacto de la percolación estándar bidimensional. Esta ecuación surge
de la hipótesis de escaleo para la longitud de correlación $\xi\propto\mid P_{s}-P_{s,c}\mid^{-\nu}$\citep{Stauffer1994}.
Para $P_{s}\rightarrow P_{s,c}$\ tenemos que $\xi\rightarrow L$,
con lo cual $L\propto\mid P_{s}-P_{s,c}\mid^{-\nu}$ desde la cual
surge la Ecuación \ref{eq:ecuacion de escaleo}.

En la Figura \ref{fig:EjemploEscaleo-de-los-Ps,c-red kagome} mostramos
a modo de ejemplo, el escaleo de los $P_{s,c}(L,P_{b})$ para cuatro
valores distintos de $P_{b}$ para la red $(3.6.3.6)$, modelo $S\cup B$.
En el límite termodinámico, $L\rightarrow\infty$, la ordenada al
origen nos da $P_{s,c}^{\infty}(P_{b})$. El conjunto de los pares
ordenados $\left(P_{b},P_{s,c}^{\infty}(P_{b})\right)_{0\leq P_{b}\leq1}$\ es
lo que se conoce como \textit{curva de fase} de la red, habiendo una
para el modelo $S\cup B$ y otra para el modelo $S\cap B$ las cuales
son complementarias. 

\begin{figure}
\begin{centering}
\includegraphics[width=8cm]{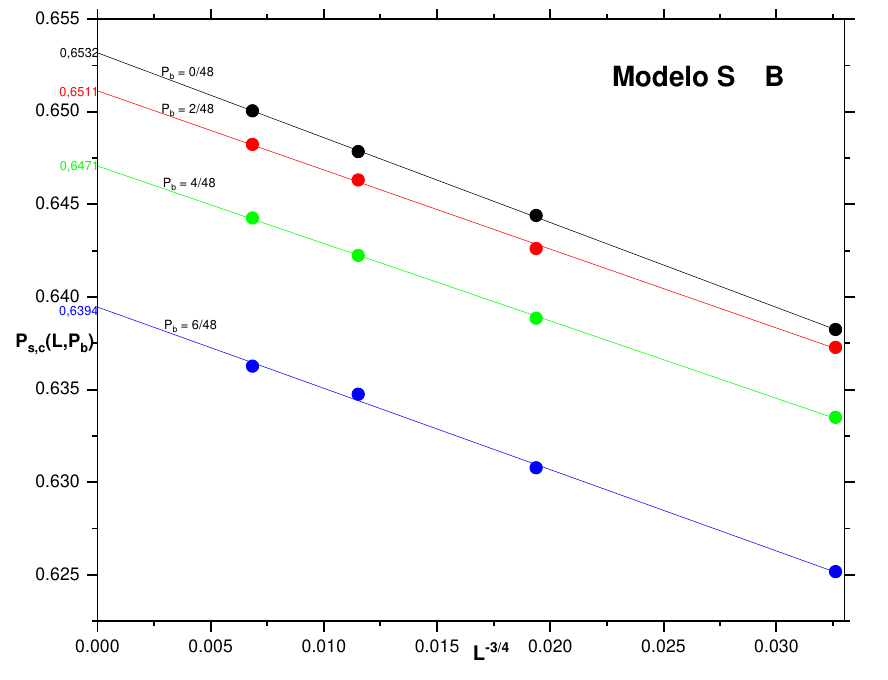}
\par\end{centering}
\caption{\label{fig:EjemploEscaleo-de-los-Ps,c-red kagome}Ejemplo de escaleo
de los $P_{s,c}(L,P_{b})$ para cuatro valores distintos de $P_{b}$,
para la red (3.6.3.6), modelo $S\cup B$.}
\end{figure}

\section{Curvas de fase: validación del proceso de obtención de las mismas}

\subsection{Preliminares}

\begin{figure}
\begin{centering}
\begin{minipage}[t]{0.5\columnwidth}%
\begin{center}
\includegraphics[width=8cm]{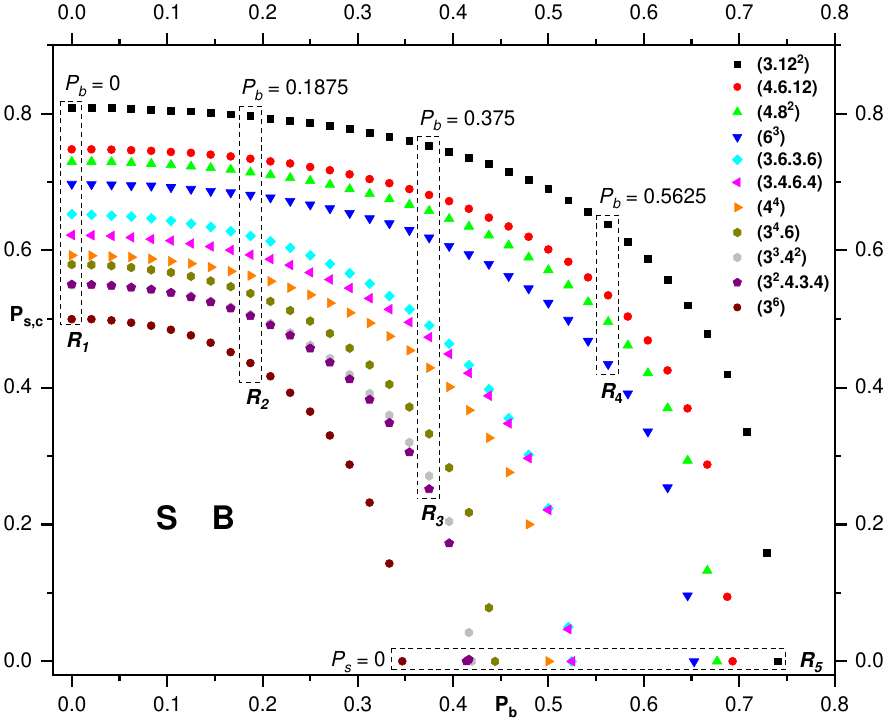}
\par\end{center}
\textbf{(a)} Curvas de fase de las redes Arquimedianas, modelo $S\cup B$.%
\end{minipage}%
\begin{minipage}[t]{0.5\columnwidth}%
\begin{center}
\includegraphics[width=8cm]{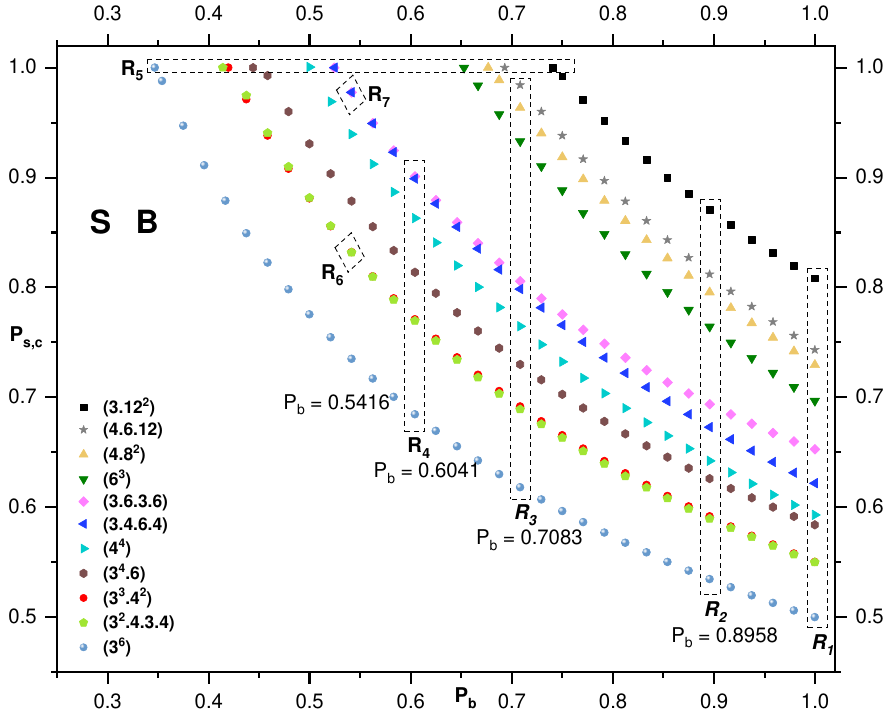}
\par\end{center}
\textbf{(b)} Curvas de fase de las redes Arquimedianas, modelo $S\cap B$.%
\end{minipage}
\par\end{centering}
\caption{\label{fig:Curvas de fase redes Arquimedianas}Curvas de fase de las
redes Arquimedianas, modelos $S\cup B$ y $S\cap B$.}
\end{figure}

En la Figura \ref{fig:Curvas de fase redes Arquimedianas} se muestran
las curvas de fase de todas las redes Arquimedianas, para ambos modelos,
obtenidas con el algoritmo descripto en las secciones anteriores. 

\subsection{Validación del proceso de simulación}

\begin{figure}
\begin{centering}
\begin{minipage}[t]{0.5\columnwidth}%
\begin{center}
\includegraphics[width=8cm]{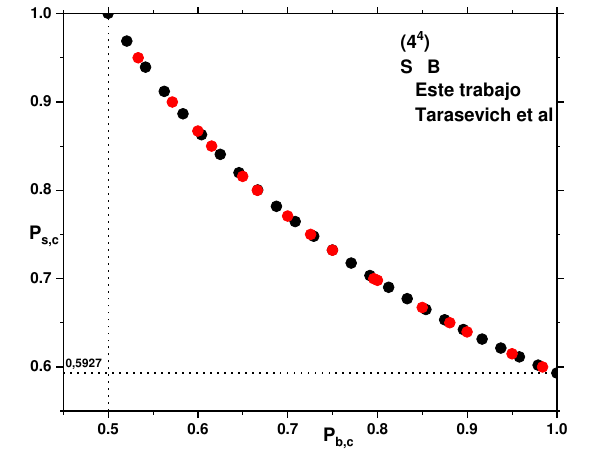}
\par\end{center}%
\end{minipage}%
\begin{minipage}[t]{0.5\columnwidth}%
\begin{center}
\includegraphics[width=8cm]{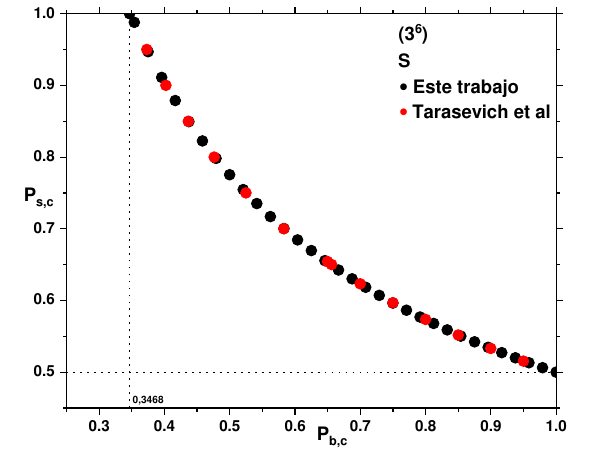}
\par\end{center}%
\end{minipage}
\par\end{centering}
\caption{\label{fig:comparacion con tarasevich-4E4-3E6}Comparación gráfica
entre los resultados numéricos obtenidos en esta tesis, con resultados
numéricos obtenidos por Tarasevich et al \citep{Tarasevich1999} para
las redes $(4^{4})$ y $(3^{6})$, respectivamente bajo el modelo
$S\cap B$.}
\end{figure}

\begin{figure}
\begin{centering}
\begin{minipage}[t]{0.5\columnwidth}%
\begin{center}
\includegraphics[width=8cm]{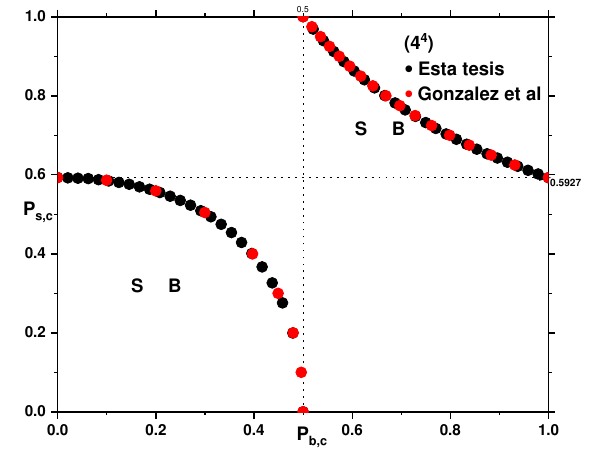}
\par\end{center}%
\end{minipage}%
\begin{minipage}[t]{0.5\columnwidth}%
\begin{center}
\includegraphics[width=8cm]{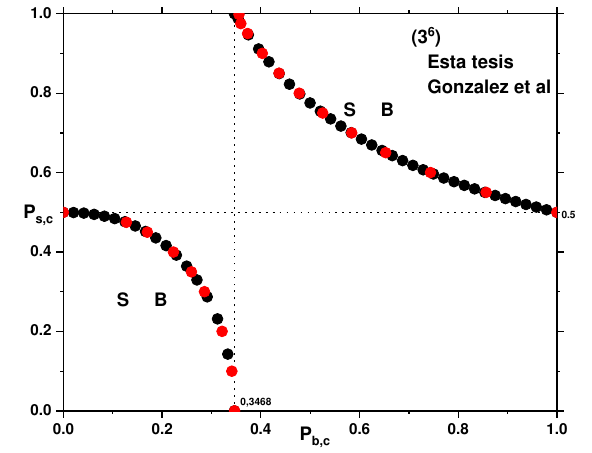}
\par\end{center}%
\end{minipage}
\par\end{centering}
\caption{\label{fig:Comparaci=0000F3n con Ramirez y Gonzalez-4E4-3E6-UNION}Comparación
gráfica entre los resultados numéricos obtenidos en esta tesis, con
resultados numéricos obtenidos por Gonzalez et al \citep{gonzalez2013site}
para las redes $(4^{4})$ y $(3^{6})$, respectivamente, bajo los
modelos $S\cap B$ y $S\cup B$.}
\end{figure}

En los gráficos expuestos en la Figura \ref{fig:comparacion con tarasevich-4E4-3E6},
hasta la Figura \ref{fig:Comparaci=0000F3n con Ramirez y Gonzalez-4E4-3E6-UNION},
se comparan los resultados numéricos obtenidos en este trabajo, a
partir del algoritmo de simulación mostrado en secciones anteriores,
con resultados de simulaciones realizadas por Tarasevich et al \citep{Tarasevich1999},
y Gonzalez et al \citep{gonzalez2013site}. Aún cuando el conjunto
usado de concentraciones de enlaces de los distintos autores, no es
coincidente con el usado en nuestro trabajo, se observa claramente
la compatibilidad gráfica de los resultados numéricos de nuestras
simulaciones, con los resultados numéricos obtenidos por ellos.

\begin{figure}
\begin{centering}
\begin{minipage}[t]{0.5\columnwidth}%
\begin{center}
\includegraphics[width=8cm]{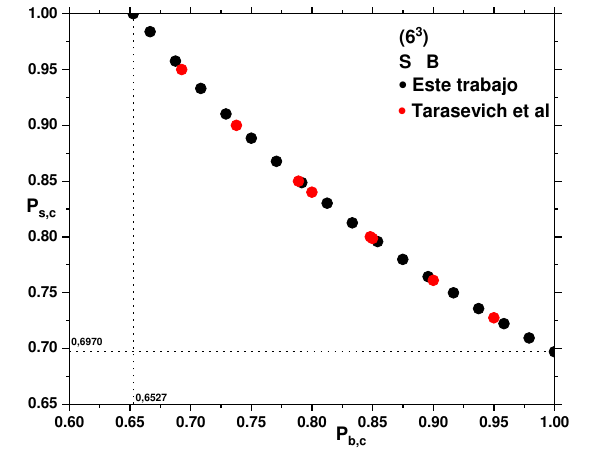}
\par\end{center}%
\end{minipage}%
\begin{minipage}[t]{0.5\columnwidth}%
\begin{center}
\includegraphics[width=8cm]{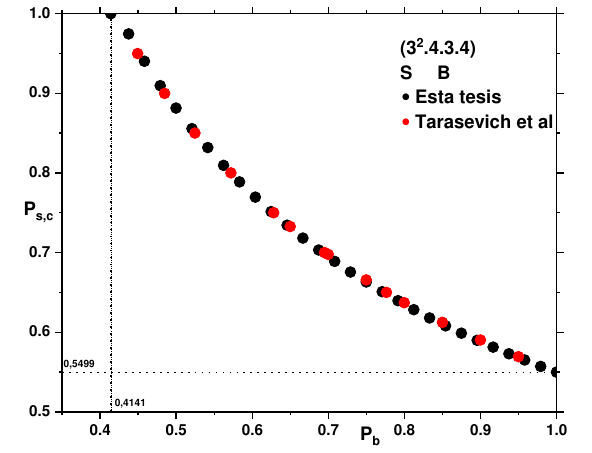}
\par\end{center}%
\end{minipage}
\par\end{centering}
\caption{\label{fig:comparacion con tarasevich-6E3-3E2C4C3C4}Comparación gráfica
entre los resultados numéricos obtenidos en esta tesis, con resultados
numéricos obtenidos por Tarasevich et al \citep{Tarasevich1999} para
las redes $(6^{3})$ y $(3^{2}.4.3.4)$, respectivamente, bajo el
modelo $S\cap B$.}
\end{figure}

\begin{figure}
\begin{centering}
\begin{minipage}[t]{0.5\columnwidth}%
\begin{center}
\includegraphics[width=8cm]{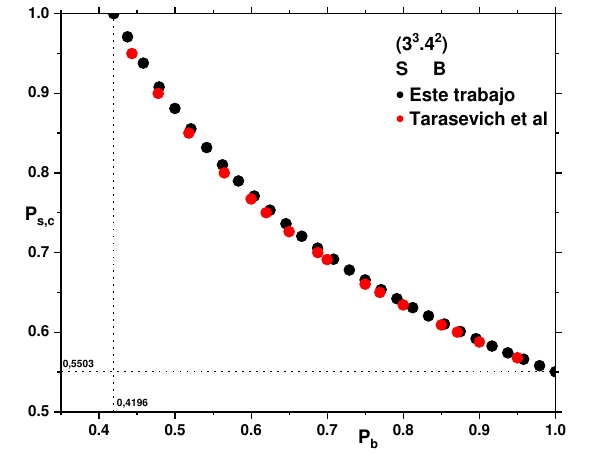}
\par\end{center}%
\end{minipage}%
\begin{minipage}[t]{0.5\columnwidth}%
\begin{center}
\includegraphics[width=8cm]{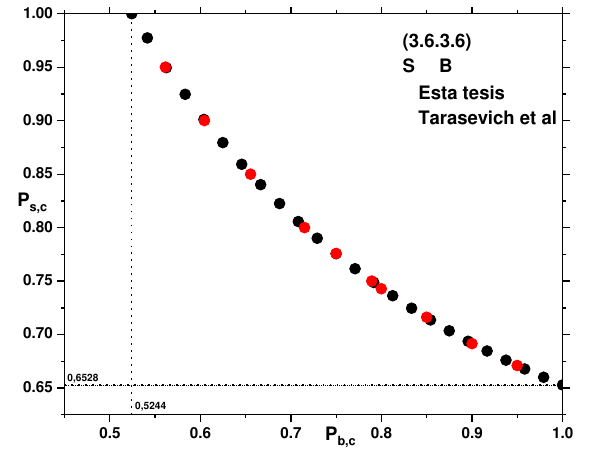}
\par\end{center}%
\end{minipage}
\par\end{centering}
\caption{\label{fig:comparacion con tarasevich-3E3C4E2-3C6C3C=000026}Comparación
gráfica entre los resultados numéricos obtenidos en esta tesis, con
resultados numéricos obtenidos por Tarasevich et al \citep{Tarasevich1999}
para las redes $(3^{3}.4^{2})$ y $(3.6.3.6)$, respectivamente, bajo
el modelo $S\cap B$.}
\end{figure}

\section{análisis de las curvas de fase del modelo $S\cup B$}

\subsection{Introducción}

En la Figura \ref{fig:Curvas de fase redes Arquimedianas}\textbf{(a)}
se muestran las curvas de fase $(P_{b},P_{s,c}(P_{b}))$ de todas
las redes Arquimedianas, para el modelo $S\cup B$, que luego describiremos.

Un grafo G1 se considera incluido en un grafo G2, si se demuestra
que G1 es un subgrafo de G2. Al ser las redes Arquimedianas grafos
planares, cada una tiene un grafo dual también planar. La red cuadrada
es auto dual, y las redes triangular y hexagonal son grafos duales
entre si. Las restantes 8 redes Arquimedianas tienen grafos duales
que no son Arquimedianas, llamadas redes de Laves . 

Parviainen R. et al\citep{parviainen2003inclusions}, encontraron
relaciones de inclusión como la establecida en el párrafo anterior,
para cada par de grafos pertenecientes al conjunto formado por las
11 redes Arquimedianas, y las 8 redes de Laves. Así, ellos determinaron
el ordenamiento parcial total por inclusión de los 19 grafos periódicos
de tamaño infinito, resumiendo sus resultados en la forma de un diagrama
de Hasse \citep{parviainen2003inclusions}. El interés de los autores
de este trabajo en el ordenamiento por inclusión de las redes, fue
motivado por el hecho de que las \textit{probabilidades críticas}
para el problema de percolación pura de sitios (enlaces), las \textit{constantes
de tiempo} en el problema de percolación de primer paso, o bien las
\textit{constantes de conectividad} en el problema de caminatas auto
evitantes, también se ordenan según como lo hagan las redes en el
ordenamiento por inclusión. Si H es un subgrafo de G, las probabilidades
críticas y la constante de tiempo, son mayores para H que para G,
y la constante de conectividad es menor para H que para G. Por lo
tanto, el conocimiento del orden por inclusión de los subgrafos permite
el uso de valores exactos o cotas conocidas de alguno de los tres
parámetros antes mencionados, para ciertos grafos o redes, como medio
de obtención de cotas para el mismo parámetro, para otros grafos o
redes. Nosotros en este trabajo solo nos focalizamos en un análisis
de inclusión que solo implica a las redes Arquimedianas, como veremos
luego en la siguiente sección.

Por otra parte, En un trabajo publicado por John C. Wierman \citep{wierman2003pairs}
el autor hizo notar que en las estimaciones de las probabilidades
críticas de percolación pura de sitios y enlaces, las cuales constaban
en la literatura de la física del momento (2001), solo había un par
de grafos o redes para el que las estimaciones de sus probabilidades
críticas de percolación pura de sitios y enlaces estaban en orden
opuestos. Si $G$ y $H$ son dos grafos, lo que casi siempre se observaba
era que:\\
Si $\mathbf{P_{s,c}^{G}>P_{s,c}^{H}}$ entonces $\mathbf{P_{b,c}^{G}>P_{b,c}^{H}}$\\
La única excepción a la regla la producían la red triangular y la
red del diamante:\\
$\mathbf{P_{s,c}^{Tri\acute{a}ngular}=0.5>P_{s,c}^{Diamante}\approx0.4299}$
pero $\mathbf{P_{b,c}^{Tri\acute{a}ngular}\approx0.3473<P_{b,c}^{Diamante}\approx0.3886}$\\
John C. Wierman se ocupó de probar en su trabajo \citep{wierman2003pairs}
que estas excepciones a la regla podían ocurrir. Lo hizo diseñando
un proceso de construcción de nuevas redes, con el cual podía generar
una colección infinita de grafos de dimensión $d\geq2$ con esta propiedad
excepcional. Ahora bien, la excepción observada por Wierman para el
par conformado por las redes triangular y diamante, puede ser también
observada para el par conformado por las redes $(3.6.3.6)$ y $(3.4.6.4)$,
y para el par de redes $(3^{3}.4^{2})$ y $(3^{2}.4.3.4)$, como veremos
en una sección posterior. Sin la luz provista por una generalización
del modelo de percolación pura sitios y enlaces, tal como lo constituyen
los modelos $S\cup B$ y $S\cap B$, es difícil visualizar las razones
de estas excepciones. Como veremos luego, estas excepciones son consecuencia
del cruce entre las curvas de fase de las redes correspondientes.
El cruce entre las curvas de fase, hace natural la inversión de las
desigualdades observadas, al pasar de la región de percolación pura
de sitios, a la región de percolación pura de enlaces, siendo ambas
regiones dibujada por el conjunto de las curvas de fase de las redes
Arquimedianas.

\subsection{Inclusiones entre redes Arquimedianas y nuevas CU con solo tres direcciones
de enlaces}

\begin{figure}
\centering{}\includegraphics[height=7cm]{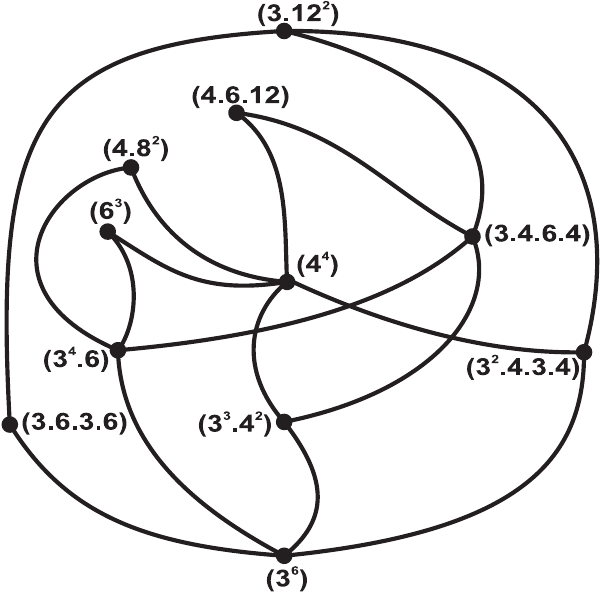}\caption{\label{fig:DiagramaHasse-RA-1-1}Diagrama de Hasse del ordenamiento
por inclusión de las $11$ redes Arquimedianas. Las líneas negras
son enlaces, los cuales representan relaciones de cobertura o inclusión
entre pares de redes Arquimedianas. Si la red H, está por encima de
otra red G en el diagrama, significa que H está incluida en G, o bien
que H es un subgrafo de G, siempre que ambas estén unidas por un enlace.}
\end{figure}

\begin{figure}
\centering{}%
\begin{minipage}[t]{0.15\columnwidth}%
\begin{center}
\includegraphics{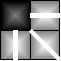}
\par\end{center}
\begin{center}
$(3^{6})$
\par\end{center}%
\end{minipage}%
\begin{minipage}[t]{0.15\columnwidth}%
\begin{center}
\includegraphics{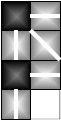}
\par\end{center}
\begin{center}
$(3^{3}.4^{2})$
\par\end{center}%
\end{minipage}%
\begin{minipage}[t]{0.25\columnwidth}%
\begin{center}
\includegraphics{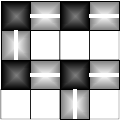}
\par\end{center}
\begin{center}
$(6^{3})$
\par\end{center}%
\end{minipage}%
\begin{minipage}[t]{0.25\columnwidth}%
\begin{center}
\includegraphics{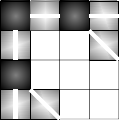}
\par\end{center}
\begin{center}
$(3.6.3.6)$
\par\end{center}%
\end{minipage}%
\begin{minipage}[t]{0.15\columnwidth}%
\begin{center}
\includegraphics{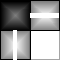}
\par\end{center}
\begin{center}
$(4^{4})$
\par\end{center}%
\end{minipage}\caption{\label{fig:CU-Alter-1-en-3E6}Cinco de las 11 CU de las redes Arquimedianas.
A lo sumo utilizan tres direcciones de enlaces para su construcción:
0, 90, y 135 grados. Las mismas utilizadas para la red triangular.Todas
las redes Arquimedianas se construyen trasladando horizontalmente
y verticalmente las respectivas CU.}
\end{figure}

\begin{figure}
\begin{centering}
\begin{minipage}[t]{0.4\columnwidth}%
\begin{center}
\includegraphics{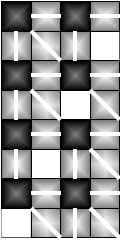}
\par\end{center}
\begin{center}
$(3^{2}.4.3.4)$
\par\end{center}%
\end{minipage}%
\begin{minipage}[t]{0.35\columnwidth}%
\begin{center}
\includegraphics{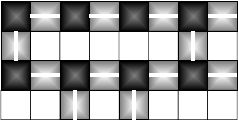}
\par\end{center}
\begin{center}
$(4.8^{2})$
\par\end{center}%
\end{minipage}%
\begin{minipage}[t]{0.2\columnwidth}%
\begin{center}
\includegraphics{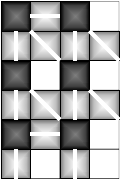}
\par\end{center}
\begin{center}
$(3.4.6.4)$
\par\end{center}%
\end{minipage}
\par\end{centering}
\begin{centering}
\noindent\begin{minipage}[t]{1\columnwidth}%
\begin{center}
\includegraphics{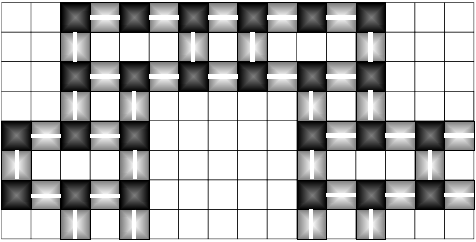}
\par\end{center}
\begin{center}
$(4.6.12)$
\par\end{center}%
\end{minipage}\caption{\label{fig:CU-Alter-3-en-3E6}Cuatro de las 11 CU de las redes Arquimedianas.
A lo sumo utilizan tres direcciones de enlaces para su construcción:
0, 90, y 135 grados. Las mismas utilizadas para la red triangular.Todas
las redes Arquimedianas se construyen trasladando horizontalmente
y verticalmente las respectivas CU.}
\par\end{centering}
\end{figure}

En este trabajo, como corolario del diagrama de Hasse presentado por
Parviainen R. et al \citep{parviainen2003inclusions}, reproducimos
en forma algo distinta, la porción del mismo que solo incluye las
relaciones de inclusión parcial y total entre las redes Arquimedianas,
tal como se muestra en La Figura \ref{fig:DiagramaHasse-RA-1-1}.
Algunas observaciones acerca de este ordenamiento parcial por inclusión
de las redes Arquimedianas: 
\begin{enumerate}
\item De las 11 redes, $(6^{3})$, $(4.8^{2})$, $(4.6.12)$ y $(3.12^{2})$
son elementos maximales. Es decir, que si H es cualquiera de los elementos
maximales, y este es un subgrafo de G, entonces no hay otro subgrafo
H\textasciiacute{} de G tal que H esté contenido en H\textasciiacute . 
\item El tamaño de la cadena mas grande (redes unidas por enlaces) es $4$,
habiendo $10$ cadenas distintas de ese tamaño. Por ejemplo la cadena
formada por las redes $(3.12^{2})$, $(3.4.6.4)$, $(3^{3}.4^{2})$,
y $(3^{6})$. 
\item La anticadena mas grande (redes no unidas entre si por enlaces) es
de tamaño $4$ y hay varias de ellas. Por ejemplo la compuesta por
las redes $(3.12^{2})$, $(4.6.12)$, $(4.8^{2})$, y $(6^{3})$.
\item Se necesitan 4 cuatro cadenas (no señaladas aquí), para que el resultado
de la unión de ellas sea el conjunto parcialmente ordenado compuesto
por las $11$ redes Arquimedianas. Por el Teorema de Dilworth \citep{dilworth1987decomposition}
el tamaño de la anticadena más grande es $4$, como fue mencionado
en el ítem anterior.
\item Todas las redes Arquimedianas están incluidas en la red triangular
$(3^{6})$. En las Figuras \ref{fig:CU-tradi-1}, y \ref{fig:CU-Alter-1},
mostramos la representación tradicional y alternativa de la celda
unidad (CU) de esta red; como se observa, ambas representaciones utilizan
tres direcciones de enlaces distintas para su construcción. Por ende,
para la construcción de todas las restantes redes Arquimedianas, solo
necesitamos tres direcciones de enlaces distintas, las mismas que
utilicemos para la red triangular. 
\item En las Figuras que van desde la Figura \ref{fig:CU-Alter-1-en-3E6},
hasta la Figura \ref{fig:CU-Alter-3-en-3E6}, mostramos las CU alternativas
de cada una de las redes Arquimedianas, para una elección particular
de las tres direcciones de enlaces utilizados para la red triangular,
a saber 0, 90, y 135 grados medidos desde la horizontal. El mínimo
común múltiplo (MCM) de todos los tamaños de aristas implicados en
las distintas CU es ahora 336. Es decir, en un cuadriculado de aristas
de tamaño $336$ o múltiplos de este número, cualquiera de las CU
de las respectivas redes Arquimedianas, cabe un número entero de veces.
Si descartamos la CU de la red $(3^{4},6)$, de tamaño $14\times14$,
el $MCM=48$. Además, todas las redes Arquimedianas pueden construirse
trasladando horizontalmente y verticalmente su respectiva CU.
\item Similar a lo dicho en la Sección \ref{subsec:CU-alternativas}, para
el caso de CU alternativas construidas con cuatro direcciones de enlaces
distintas, cada sitio y enlace de una red dada, puede etiquetarse
con un par ordenado de números enteros, y ser caracterizado mediante
la función módulo. Por otra parte, a igual tamaño de red, el conjunto
de todos los sitios y enlaces que definen una red dada, pueden ser
etiquetados y caracterizados por el mismo conjunto o subconjunto de
pares ordenados utilizados para etiquetar y caracterizar los sitios
y enlaces de la red triangular. 
\item En función de lo observado en los ítems 5, 6 y 7, es pertinente afirmar
que la red triangular (3\textsuperscript{{\tiny{}6}}) constituye
una red o plataforma geométrica base para el estudio de las restantes
redes Arquimedianas. Dada una propiedad medible en esta, también será
medible en cualquiera de las restantes redes Arquimedianas y estas
mediciones se podrán comparar. El valor medido de la propiedad en
la red triangular, constituirá una unidad de medida de la misma para
el resto de las redes Arquimedianas. 
\end{enumerate}

\begin{figure}
\begin{centering}
\begin{minipage}[t]{0.5\columnwidth}%
\begin{center}
\includegraphics{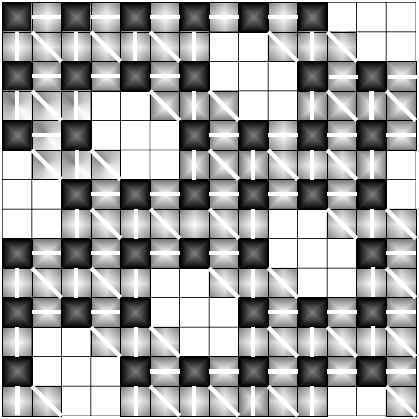}
\par\end{center}
\begin{center}
$(3^{4}.6)$
\par\end{center}%
\end{minipage}%
\begin{minipage}[t]{0.5\columnwidth}%
\begin{center}
\includegraphics[scale=0.9]{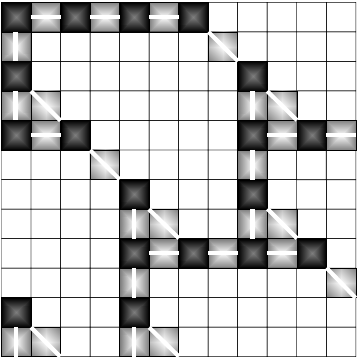}
\par\end{center}
\begin{center}
$(3.12^{2})$
\par\end{center}%
\end{minipage}
\par\end{centering}
\caption{\label{fig:CU-Alter-2-en-3E6}Tres de las 11 CU de las redes Arquimedianas.
A lo sumo utilizan tres direcciones de enlaces para su construcción:
0, 90, y 135 grados. Las mismas utilizadas para la red triangular.
Todas las redes Arquimedianas se construyen trasladando horizontalmente
y verticalmente las respectivas CU.}
\end{figure}

\subsection{Ordenamiento por inclusión de las curvas de fase del modelo $S\cup B$
e inversión}

Tal como se observa en la Figura \ref{fig:Curvas de fase redes Arquimedianas}\textbf{(a)},
las curvas de fase de la redes Arquimedianas en el modelo $S\cup B$,
se ordenan según el criterio inducido por el diagrama de fase mostrado
en la Figura \ref{fig:DiagramaHasse-RA-1-1}. A continuación detallamos
esta observación:

\subsubsection{Ordenamiento por inclusión en los rectángulos $R_{1}$ y $R_{5}$}

\noindent \textbf{Rectángulo $R_{1}$:} Contiene los puntos $(0,P_{s,c}(0))$
de las distintas redes Arquimedianas. Estos puntos corresponden a
los valores críticos de percolación pura de sitios. En la Figura \ref{fig:Ordenamiento-de-R1-R5-RedesArqui-union}\textbf{(b)}
y \textbf{(c)}, se muestra el diagrama de Hasse del ordenamiento de
mayor a menor de los valores críticos de percolación pura de sitios,
$P_{s,c}(0)$.

\noindent \textbf{Rectángulo $R_{5}$:} Contiene los puntos $(P_{b,c}(0),0)$
de las distintas redes Arquimedianas. Estos puntos corresponden a
los valores críticos de percolación pura de enlaces. En la Figura
\ref{fig:Ordenamiento-de-R1-R5-RedesArqui-union}\textbf{(e)} y \textbf{(f)},
se muestra el diagrama de Hasse del ordenamiento de mayor a menor
de los valores críticos de percolación pura de enlaces, $P_{b,c}(0)$.
En las Figuras \ref{fig:Ordenamiento-de-R1-R5-RedesArqui-union}\textbf{(a)}
y \textbf{(g)}, se muestra el diagrama de Hasse del ordenamiento parcial
y total por inclusión de la redes Arquimedianas. Ambos diagramas representan
el mismo diagrama de Hasse que el de la Figura \ref{fig:DiagramaHasse-RA-1-1}. 

\subsubsection{Ordenamiento por inclusión en los rectángulos $R_{2}$-$R_{3}$-$R_{4}$}

\noindent \textbf{Rectángulos} $R_{2}$, $R_{3},$ R$_{4}$: Contienen
los puntos $(0.1875,P_{s,c}(0.1875))$, $(0.375,P_{s,c}(0.375))$,
y $(0.5625,P_{s,c}(0.5625))$, respectivamente, para las distintas
redes Arquimedianas.

\begin{figure}
\begin{centering}
\includegraphics{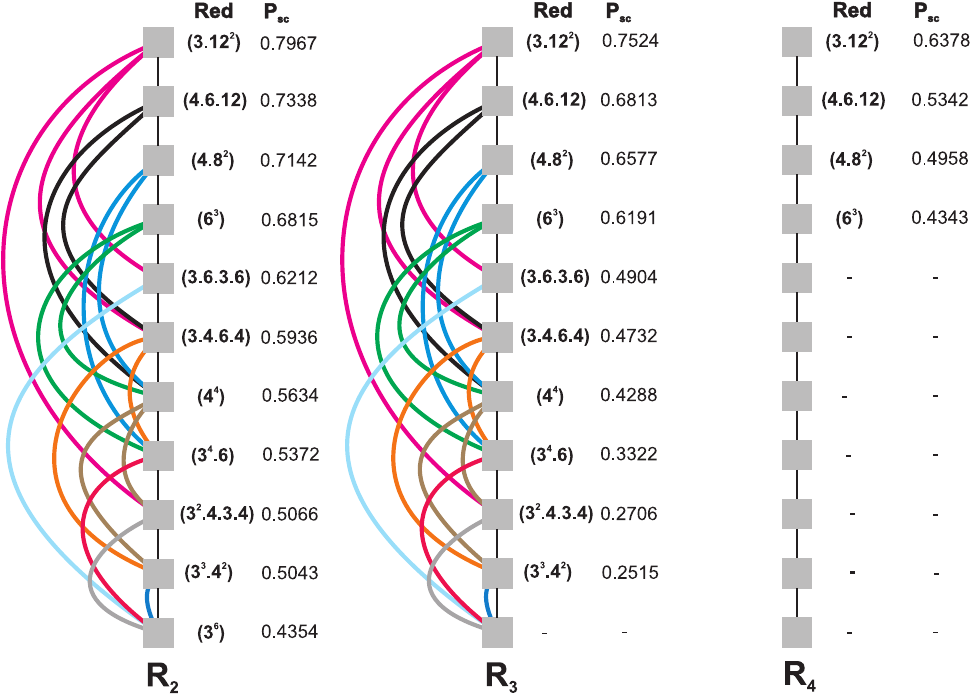}
\par\end{centering}
\caption{\label{fig:Ordenamiento-por-inclusi=0000F3n-y-valores-criticos-en R2-R3-R4-2-union}Ordenamiento
por inclusión parcial y total, y por valores críticos de percolación
sitio-enlace de los puntos de las curvas de fase contenidos en los
rectángulos $R_{2}$, $R_{3}$, y $R_{4}$ de la Figura \ref{fig:Curvas de fase redes Arquimedianas}\textbf{(a)}.}
\end{figure}

En la Figura \ref{fig:Ordenamiento-por-inclusi=0000F3n-y-valores-criticos-en R2-R3-R4-2-union}
mostramos el ordenamiento por inclusión parcial y total, y por valores
críticos de percolación sitio-enlace de los puntos de las curvas de
fase contenidos en los rectángulos $R_{2}$, $R_{3}$, y $R_{4}$
de la Figura \ref{fig:Curvas de fase redes Arquimedianas}\textbf{(a)}.
Algunas consideraciones sobre estos ordenamientos:
\begin{enumerate}
\item El ordenamiento asociado a $R_{2}$, es también válido para cualquier
rectángulo vertical similar a este, que contenga los puntos $P_{s,c}(P_{b})$
sobre las distintas curvas de fases, con $P_{b}$ fijo, tal que $0<P_{b}\leq0.3468$,
siendo la cota superior de este semi intervalo, el valor crítico de
percolación de enlaces puro para la red $(3^{6})$.
\item El ordenamiento asociado a $R_{3}$, en donde ya no aparece la red
$(3^{6})$, es también válido para cualquier rectángulo vertical similar
a este que, contenga los puntos $P_{s,c}(P_{b})$ sobre las distintas
curvas de fases, con $P_{b}$ fijo, tal que $0.3468<P_{b}\leq0.4167$,
siendo la cota superior de este semi intervalo, el valor crítico de
percolación de enlaces puro para la red $(3^{2}.4.3.4)$. 
\item Podemos realizar ordenamientos similares al realizado para $R_{3}$,
válidos para otros semi intervalos de $P_{b}$. Estos ordenamientos
implicarán cada vez menos redes.
\item El ordenamiento asociado a $R_{4}$, en donde solo aparecen las redes
$(3.12^{2})$, $(4.6.12)$, $(4.8^{2})$, y $(6^{3})$, es también
válido para cualquier rectángulo vertical similar a este que, contenga
los puntos $P_{s,c}(P_{b})$ sobre las distintas curvas de fases,
con $P_{b}$ fijo, tal que $0.5244<P_{b}\leq0.6527$, siendo la cota
inferior de este semi intervalo el valor crítico de percolación de
enlaces puros para la red $(3.6.3.6)$, y la cota superior de este
semi intervalo, el valor crítico de percolación de enlaces puro para
la red $(6^{3})$. 
\item El último ordenamiento posible, en donde solo aparecería la red $(3.12^{2})$,
es el que se da en el intervalo $0.6931<P_{b}\leq0.7404$, siendo
la cota inferior de este semi intervalo el valor crítico de percolación
de enlaces puros para la red $(4.6.12)$, y la cota superior de este
semi intervalo, el valor crítico de percolación de enlaces puro para
la red $(3.12^{2})$. 
\end{enumerate}

\subsubsection{Inversiones }

\noindent En la Figura \ref{fig:DiagramaHasse-RA-1-1}\textbf{(d)},
Señalamos la inversión de valores entre los umbrales críticos de percolación
pura de sitios y de enlaces para el par de redes $(3.6.3.6)$ y $(3.4.6.4)$
y para el par $(3^{2}.4.3.4)$ , $(3^{3}.4^{2})$. Como veremos luego,
esta inversión se debe al cruce entre las curvas de fases que conforman
cada par, y no contradicen lo predicho por las relaciones de inclusión
pues las redes que conforman cada par forman o son parte de una anticadena.\\
\textbf{Inversión entre} $(3.6.3.6)$ y $(3.4.6.4)$:\\
 $P_{s,c}^{(3.6.3.6)}(0)>P_{s,c}^{(3.4.6.4)}(0)$. Esta desigualdad
se mantiene para $P_{b}>0$ hasta el valor $P_{b}^{I}$ perteneciente
al intervalo $0.5208<P_{b}^{I}<0.5248$, donde $0.5208$ es la coordenada
horizontal del punto $(0.5208,P_{s,c}(0.5208))$ sobre la curva de
fase de la red $(3.6.3.6)$ o bien de la red $(3.4.6.4)$, y $0.5248=P_{b,c}^{(3.4.6.4)}(0)$.
Para $P_{b}>P_{b}^{I}$ la desigualdad se invierte, es decir, $P_{s,c}^{(3.6.3.6)}(P_{b})<P_{s,c}^{(3.4.6.4)}(P_{b})$.
Esta inversión hace que se verifique esta otra desigualdad, en el
ordenamiento de mayor a menor de los valores críticos de percolación
pura de enlaces:\\
 $P_{b,c}^{(3.6.3.6)}(0)<P_{b,c}^{(3.4.6.4)}(0)$. \\
\textbf{Inversión entre} $(3^{3}.4^{2})$ y $(3^{2}.4.3.4)$:\\
 $P_{s,c}^{(3^{2}.4.3.4)}(0)>P_{s,c}^{(3^{3}.4^{2})}(0)$. Esta desigualdad
se mantiene mientras la concentración de enlaces verifique la desigualdad
$P_{b}<0.021$. En $P_{b}\approx0.021$ ambas curvas de fase se interceptan,
y para $P_{b}>0.021$ la desigualdad se invierte, es decir, $P_{s,c}^{(3^{2}.4.3.4)}(P_{b})<P_{s,c}^{(3^{3}.4^{2})}(P_{b})$.
Esta inversión hace que se verifique esta otra desigualdad, en el
ordenamiento de mayor a menor de los valores críticos de percolación
pura de enlaces:\\
$P_{b,c}^{(3^{2}.4.3.4)}(0)<P_{b,c}^{(3^{3}.4^{2})}(0)$.

\begin{figure}
\begin{centering}
\includegraphics[width=11cm]{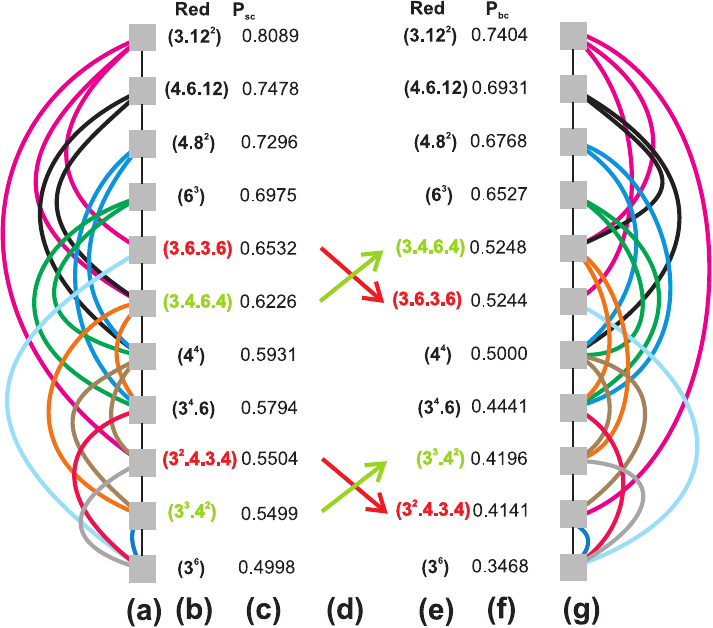}
\par\end{centering}
\caption{\textbf{\label{fig:Ordenamiento-de-R1-R5-RedesArqui-union}(a) y (g)}
Diagrama de Hasse del ordenamiento por inclusión parcial y total de
la redes Arquimedianas. \textbf{(b) }y\textbf{ (c)} Diagrama de Hasse
del ordenamiento de mayor a menor de los valores críticos de percolación
pura de sitios, $P_{s,c}(0)$.\textbf{ (d)} Señalamiento del intercambio
de posiciones entre las redes $(3.6.3.6)$ , $(3.4.6.4)$ y las redes
$(3^{2}.4.3.4)$ , $(3^{3}.4^{2})$. \textbf{(e)} y \textbf{(f)} Diagrama
de Hasse del ordenamiento de mayor a menor de valores críticos de
percolación pura de enlaces, $P_{b,c}(0)$.}
\end{figure}

\subsection{Ordenamiento por inclusión de las curvas de fase del modelo $S\cap B$
e inversión}

Tal como se observa en la Figura \ref{fig:Curvas de fase redes Arquimedianas}\textbf{(b)},
las curvas de fase de la redes Arquimedianas en el modelo $S\cap B$,
se ordenan según el criterio inducido por el diagrama de fase mostrado
en la Figura \ref{fig:DiagramaHasse-RA-1-1}. A continuación detallamos
esta observación:

\subsubsection{Ordenamiento por inclusión en los rectángulos $R_{1}$ y $R_{5}$}

\noindent \textbf{Rectángulo $R_{1}$:} Contiene los puntos $(1,P_{s,c}(1))$
de las distintas redes Arquimedianas. Estos puntos corresponden a
los valores críticos de percolación pura de sitios. En la Figura \ref{fig:Ordenamiento-de-R1-R5-RedesArqui-interseccion}\textbf{(b)}
y \textbf{(c)}, se muestra el diagrama de Hasse del ordenamiento de
mayor a menor de los valores críticos de percolación pura de sitios,
$P_{s,c}(1)$.

\noindent \textbf{Rectángulo $R_{5}$:} Contiene los puntos $(P_{b,c}(1),1)$
de las distintas redes Arquimedianas. Estos puntos corresponden a
los valores críticos de percolación pura de enlaces. En la Figura
\ref{fig:Ordenamiento-de-R1-R5-RedesArqui-interseccion}\textbf{(e)}
y \textbf{(f)}, se muestra el diagrama de Hasse del ordenamiento de
mayor a menor de los valores críticos de percolación pura de enlaces,
$P_{b,c}(1)$. En las Figuras \ref{fig:Ordenamiento-de-R1-R5-RedesArqui-interseccion}\textbf{(a)}
y \textbf{(g)}, se muestra el diagrama de Hasse del ordenamiento parcial
y total por inclusión de la redes Arquimedianas. Ambos diagramas representan
el mismo diagrama de Hasse que el de la Figura \ref{fig:DiagramaHasse-RA-1-1}. 

\subsubsection{Ordenamiento por inclusión en los rectángulos $R_{2}$-$R_{3}$-$R_{4}$}

\noindent \textbf{Rectángulos} $R_{2}$, $R_{3},$ R$_{4}$: Contienen
los puntos $(0.8953,P_{s,c}(0.8953))$ , $(0.7083,P_{s,c}(0.7083))$
, y $(0.6042,P_{s,c}(0.6042))$, respectivamente, para las distintas
redes Arquimedianas.

\begin{figure}
\begin{centering}
\includegraphics[width=12cm]{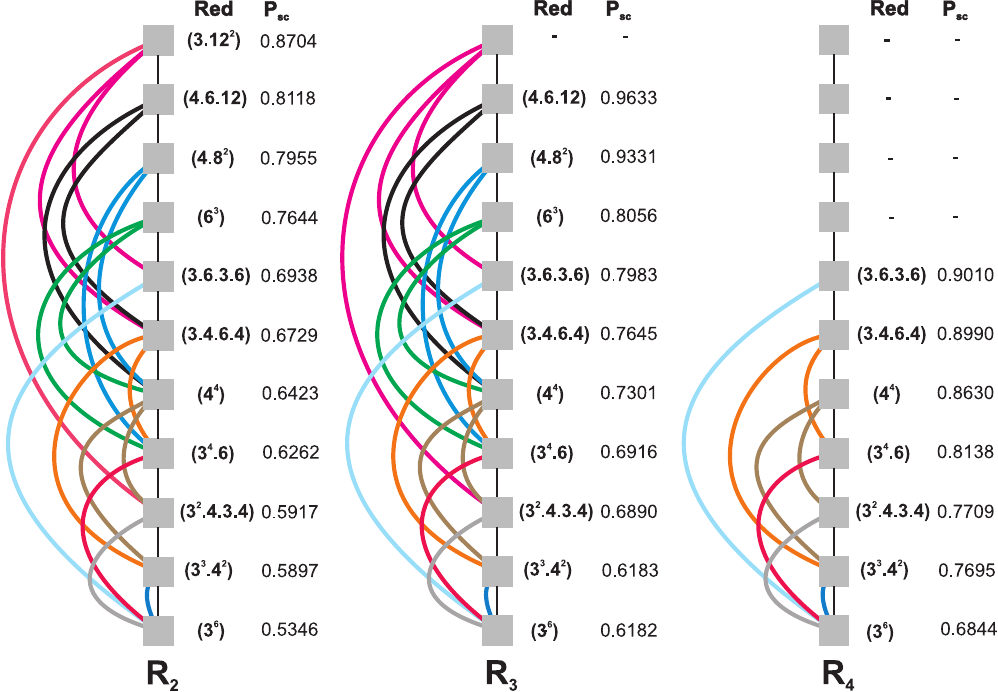}
\par\end{centering}
\caption{\label{fig:Ordenamiento-por-inclusi=0000F3n-y-valores-criticos-en R2-R3-R4-interseccion}Ordenamiento
por inclusión parcial y total, y por valores críticos de percolación
sitio-enlace de los puntos de las curvas de fase contenidos en los
rectángulos $R_{2}$, $R_{3}$, y $R_{4}$ de la Figura \ref{fig:Curvas de fase redes Arquimedianas}\textbf{(b)}.}
\end{figure}

En la Figura \ref{fig:Ordenamiento-por-inclusi=0000F3n-y-valores-criticos-en R2-R3-R4-interseccion}
mostramos el ordenamiento por inclusión parcial y total, y por valores
críticos de percolación sitio-enlace de los puntos de las curvas de
fase contenidos en los rectángulos $R_{2}$, $R_{3}$, y $R_{4}$
de la Figura \ref{fig:Curvas de fase redes Arquimedianas}\textbf{(b)}.
Algunas consideraciones sobre estos ordenamientos:
\begin{enumerate}
\item El ordenamiento asociado a $R_{2}$, es también válido para cualquier
rectángulo vertical similar a este que, contenga los puntos $P_{s,c}(P_{b})$
sobre las distintas curvas de fases, con $P_{b}$ fijo, tal que $0.7404\leq P_{b}<0$,
siendo la cota inferior de este semi intervalo, el valor crítico de
percolación de enlaces puro para la red $(3.12^{2})$.
\item El ordenamiento asociado a $R_{3}$, en donde ya no aparece la red
$(3.12^{2})$, es también válido para cualquier rectángulo vertical
similar a este que, contenga los puntos $P_{s,c}(P_{b})$ sobre las
distintas curvas de fases, con $P_{b}$ fijo, tal que $0.6931\leq P_{b}<0.7404$,
siendo la cota inferior de este semi intervalo, el valor crítico de
percolación de enlaces puro para la red $(4.6.12)$. 
\item Podemos realizar ordenamientos similares al realizado para $R_{3}$,
válidos para otros semi intervalos de $P_{b}$. Estos ordenamientos
implicarán cada vez menos redes.
\item El ordenamiento asociado a $R_{4}$, es también válido para cualquier
rectángulo vertical similar a este que, contenga los puntos $P_{s,c}(P_{b})$
sobre las distintas curvas de fases, con $P_{b}$ fijo, tal que $0.5244\leq P_{b}<0.6527$,
siendo la cota inferior de este semi intervalo el valor crítico de
percolación de enlaces puros para la red $(3.6.3.6)$, y la cota superior
de este semi intervalo, el valor crítico de percolación de enlaces
puro para la red $(6^{3})$. 
\item El último ordenamiento posible, en donde solo aparece la red $(3^{6})$,
es el que se da en el intervalo $0.3468\leq P_{b}<0.4141$, siendo
la cota inferior de este semi intervalo el valor crítico de percolación
de enlaces puros para la red $(3^{6})$, y la cota superior de este
semi intervalo, el valor crítico de percolación de enlaces puro para
la red $(3^{2}.4.3.4)$. 
\end{enumerate}

\subsubsection{Inversiones}

\begin{figure}
\begin{centering}
\includegraphics[width=11cm]{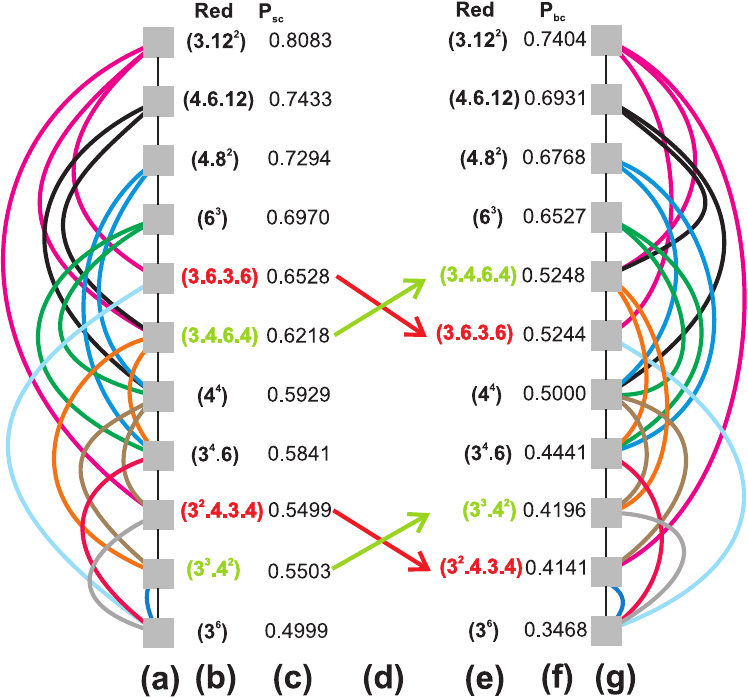}
\par\end{centering}
\caption{\textbf{\label{fig:Ordenamiento-de-R1-R5-RedesArqui-interseccion}(a)
y (g)} Diagrama de Hasse del ordenamiento por inclusión parcial y
total de la redes Arquimedianas. Diagrama de Hasse del ordenamiento
parcial y total por inclusión de la redes Arquimedianas. \textbf{(b)
}y\textbf{ (c)} Diagrama de Hasse del ordenamiento de mayor a menor
de los valores críticos de percolación pura de sitios, $P_{s,c}(1)$.\textbf{
(d)} Señalamiento del intercambio de posiciones entre las redes $(3.6.3.6)$
, $(3.4.6.4)$ y las redes $(3^{2}.4.3.4)$ , $(3^{3}.4^{2})$. \textbf{(e)}
y \textbf{(f)} Diagrama de Hasse del ordenamiento de mayor a menor
de valores críticos de percolación pura de enlaces, $P_{b,c}(1)$.}
\end{figure}

En la Figura \ref{fig:DiagramaHasse-RA-1-1}\textbf{(d)}, Señalamos
la inversión de valores entre los umbrales críticos de percolación
pura de sitios y de enlaces entre el par de redes $(3.6.3.6)$ y $(3.4.6.4)$
y el par $(3^{2}.4.3.4)$ , $(3^{3}.4^{2})$. Como veremos luego,
esta inversión también se debe al cruce entre las curvas de fases
que conforman cada par, y no contradicen lo predicho por las relaciones
de inclusión pues las redes que conforman cada par forman o son parte
de una anticadena.\\
\textbf{Inversión entre} $(3.6.3.6)$ y $(3.4.6.4)$:\\
 $P_{s,c}^{(3.6.3.6)}(1)>P_{s,c}^{(3.4.6.4)}(1)$. Esta desigualdad
se mantiene para $P_{b}<1$ hasta aproximadamente$P_{b}^{I}=0.5417$
donde $P_{s,c}^{(3.6.3.6)}(P_{b}^{I})=P_{s,c}^{(3.4.6.4)}(P_{b}^{I})=0.9773$,
es decir, hasta el punto donde las curvas de fase se interceptan,
contenido en el rectángulo $R_{7}$. En $P_{b}^{I}$ la desigualdad
se invierte, y ahora para $P_{b}<P_{b}^{I}$ se verifica que $P_{s,c}^{(3.6.3.6)}(P_{b})<P_{s,c}^{(3.4.6.4)}(P_{b})$.
Esta inversión hará que se verifique la desigualdad:\\
 $P_{b,c}^{(3.6.3.6)}(1)<P_{b,c}^{(3.4.6.4)}(1)$.\\
Esta desigualdad, es similar a la desigualdad señalada en la Figura
\ref{fig:Ordenamiento-de-R1-R5-RedesArqui-union}, en el ordenamiento
de mayor a menor de los valores críticos de percolación pura de enlaces
$P_{b,c}$(0). También la observamos en la Figura \ref{fig:Ordenamiento-de-R1-R5-RedesArqui-interseccion},
en el ordenamiento de mayor a menor de los valores críticos de percolación
pura de enlaces $P_{b,c}$(1).\\
\textbf{Inversión entre} $(3^{3}.4^{2})$ y $(3^{2}.4.3.4)$:\\
 $P_{s,c}^{(3^{3}.4^{2})}(1)>P_{s,c}^{(3^{2}.4.3.4)}(1)$. Esta desigualdad
se mantiene para $P_{b}<1$ hasta aproximadamente $P_{b}=0.5417$donde
$P_{s,c}^{(3^{3}.4^{2})}(P_{b})=P_{s,c}^{(3^{2}.4.3.4)}(P_{b})=0.8318,$es
decir, hasta el punto donde las curvas de fase se interceptan, contenido
en el rectángulo $R_{6}$. Para $P_{b}<0.5417$ la desigualdad se
invierte, ahora $P_{s,c}^{(3^{3}.4^{2})}(P_{b})<P_{s,c}^{(3^{2}.4.3.4)}(P_{b})$
se mantendrá hasta un punto $P_{b}^{I}$ del intervalo $0.4141\leq P_{b}^{I}<0.4375$,
en donde nuevamente $P_{s,c}^{(3^{3}.4^{2})}(P_{b}^{I})=P_{s,c}^{(3^{2}.4.3.4)}(P_{b}^{I})$.
Para $P_{b}<P_{b}^{I}$se invertirá nuevamente, es decir, $P_{s,c}^{(3^{3}.4^{2})}(P_{b})>P_{s,c}^{(3^{2}.4.3.4)}(P_{b})$.
Esta última relación hará que se produzca la desigualdad:\\
$P_{b,c}^{(3^{3}.4^{2})}(1)>P_{b,c}^{(3^{2}.4.3.4)}(1)$\\
Esta desigualdad, es similar a la desigualdad señalada en la Figura
\ref{fig:Ordenamiento-de-R1-R5-RedesArqui-union}, en el ordenamiento
de mayor a menor de los valores críticos de percolación pura de enlaces
$P_{b,c}$(0). También la observamos en la Figura \ref{fig:Ordenamiento-de-R1-R5-RedesArqui-interseccion},
en el ordenamiento de mayor a menor de los valores críticos de percolación
pura de enlaces $P_{b,c}$(1).

\pagebreak{}

\section{Conclusiones}
\begin{enumerate}
\item Hemos presentado una forma geométrica alternativa de representar cada
una de las redes Arquimedianas. Esta representación se fundamenta
en la construcción de celdas unidad (CU) para cada una de ellas, las
cuales trasladadas horizontalmente y verticalmente, nos permiten alcanzar
el tamaño deseado de la red en estudio. Las CU propuestas utilizan
hasta cuatro direcciones de enlaces 0°, 45°, 90°, y 135°, respectivamente,
medidos respecto del eje horizontal; las mismas permiten que las aristas
de cada red Arquimediana, pueden ser elegidas de manera tal que sean
un múltiplo de 48, lo que permite diseñar a todas la redes Arquimedianas
del mismo tamaño, facilitando la realización de estudios de comparación
entre ellas sin preocupación por la disparidad de tamaños entre redes.
Usamos estas CU para la determinación de las curvas de fase de todas
las redes Arquimedianas.
\item En el marco de la representación alternativa propuesta, hemos complementando
el trabajo de Parviainen et al \citep{parviainen2003inclusions},
presentando una CU alternativa para cada una de las redes Arquimedianas,
las cuales utilizan a lo sumo tres direcciones de enlaces distintas
0°, 90°, y 135°, respectivamente, medidos respecto del eje horizontal;
las mismas permiten que las aristas de cada red Arquimediana, pueden
ser elegidas de manera tal que sean un múltiplo de 336, lo que también
permite diseñar a todas la redes Arquimedianas del mismo tamaño, facilitando
la realización de estudios de comparación entre ellas sin preocupación
por la disparidad de tamaños entre redes. Una cuestión importante
de estas CU en el caso de ser elegidas todas del mismo tamaño, es
que el mismo conjunto de pares ordenados de números enteros utilizados
para etiquetar y caracterizar los sitios y enlaces de la red triangular,
puede ser usado para etiquetar y caracterizar los sitios y enlaces
de cada una de las restantes redes Arquimedianas, pues estas están
en ella.
\item Hemos mostrado que la dupla \textquotedbl CU alternativa, función
módulo\textquotedbl{} permite caracterizar por completo cada sitio
y enlace de cada red Arquimediana, facilitando enormemente el tratamiento
computacional de las mismas.
\item Hemos desarrollado y puesto a consideración, un proceso computacional
que nos permite obtener las curvas de fase asociadas a la deposición
aleatoria de monómeros, para cada una de las redes Arquimedianas.
Validamos el proceso computacional con datos obtenidos por Tarasevich
et al \citep{Tarasevich1999} para las redes $(4^{4})$, $(3^{6})$,
$(6^{3})$, $(3^{3}.4^{2})$, $(3^{2}.4.3.4)$ y $(3.6.3.6)$, en
el caso del modelo $S\cap B$; y con datos obtenidos por Gonzalez
et al \citep{gonzalez2013site,gonzalez2016site} para las redes $(4^{4})$
y $(3^{6})$ en el caso de los modelos $S\cap B$ y $S\cup B$, respectivamente.
\item Presentamos por primeras vez las curvas de fase asociadas al modelo
$S\cap B$, para las redes $(3.4.6.4)$, $(4.6.12)$, $(4.8^{2})$,
$(3.12^{2})$, y $(3^{4}.6)$, respectivamente; y las curvas de fase
asociadas al modelo $S\cup B$, para las redes $(6^{3})$, $(3^{3}.4^{2})$,
$(3^{2}.4.3.4)$ , $(3.6.3.6)$, $(3.4.6.4)$, $(4.6.12)$, $(4.8^{2})$,
$(3.12^{2})$, y $(3^{4}.6)$, respectivamente.
\item Hemos mostrado que el conjunto formado por las curvas de fase por
deposición de monómeros, de cada una de las redes Arquimedianas, se
ordenan en el espació de fase tal como lo inducen las relaciones de
inclusión parcial y total entre ellas, tanto en el modelo $S\cup B$,
como en el modelo $S\cap B$. 
\item Hemos mostrado que la inversión de las desigualdades $P_{s,c}^{G}>P_{s,c}^{H}$,
entonces $P_{b,c}^{G}<P_{s,c}^{H}$, observada en este trabajo par
el par de redes $(3.6.3.6)$ y (3$.4.6.4)$ y para el par $(3^{3}.4^{2})$
y $(3^{2}.4.3.4)$, tanto en el modelo $S\cup B$, como en el modelo
$S\cap B$, se deben al cruce entre las curvas de fase de las redes
que conforman cada par. Como las redes que conforman cada par implicado
en las inversiones observadas, forman o son parte de una anticadena,
esta conclusión no contradice lo expresado en el ítem 6. 
\end{enumerate}
\clearpage

\bibliographystyle{aapmrev4-2}
\bibliography{bibliografiaRepresentacionAlternativaYCurvasdeFase}

\end{document}